\documentstyle[12pt,graphicx]{article}                
\begin{document}
%This is dvips(k) 5.86 Cobegin{document}
\baselineskip 18pt
%t
\def\today{\ifcase\month\or
 January\or February\or March\or April\or May\or June\or
 July\or August\or September\or October\or November\or December\fi
 \space\number\day, \number\year}
\def\thebibliography#1{\section*{References\markboth
 {References}{References}}\list
 {[\arabic{enumi}]}{\settowidth\labelwidth{[#1]}
 \leftmargin\labelwidth
 \advance\leftmargin\labelsep
 \usecounter{enumi}}
 \def\newblock{\hskip .11em plus .33em minus .07em}
 \sloppy
 \sfcode`\.=1000\relax}
\let\endthebibliography=\endlist
%%%%%%%%%%%%%%%%%%%%%%%%%%%%%%%%%
%%%%%%%%%%%%%%%%%%%%%%%%%%%%%
% A useful Journal macro
\def\Journal#1#2#3#4{{#1} {\bf #2}, #3 (#4)}

% Some useful journal names
\def\NCA{\em Nuovo Cimento}
\def\NIM{\em Nucl. Instrum. Methods}
\def\NIMA{{\em Nucl. Instrum. Methods} A}
\def\NPA{{\em Nucl. Phys.} A}
\def\NPB{{\em Nucl. Phys.} B}
\def\PLB{{\em Phys. Lett.}  B}
\def\PRL{\em Phys. Rev. Lett.}
\def\PRD{{\em Phys. Rev.} D}
\def\ZPC{{\em Z. Phys.} C}

%%%%%%%%%%%%%%%%%%%%%%%%%%%%
\newcommand{\alpi}{\it O(\alpha/\pi)}
\def\beq{\begin{equation}}
\def\eeq{\end{equation}}
\def\lsim{\ ^<\llap{$_\sim$}\ }
\def\gsim{\ ^>\llap{$_\sim$}\ }
\def\r2{\sqrt 2}
\def\beq{\begin{equation}}
\def\eeq{\end{equation}}
\def\beqn{\begin{eqnarray}}
\def\eeqn{\end{eqnarray}}
\def\rmuu{\gamma^{\mu}}
\def\rmud{\gamma_{\mu}}
\def\PL{{1-\gamma_5\over 2}}
\def\PR{{1+\gamma_5\over 2}}
\def\sinW2{\sin^2\theta_W}
\def\AEM{\alpha_{EM}}
\def\mul{M_{\tilde{u} L}^2}
\def\mur{M_{\tilde{u} R}^2}
\def\mdl{M_{\tilde{d} L}^2}
\def\mdr{M_{\tilde{d} R}^2}
\def\mz2{M_{z}^2}
\def\c2b{\cos 2\beta}
\def\au{A_u}
\def\ad{A_d}
\def\cob{\cot \beta}
\def\v#1{v_#1}
\def\tb{\tan\beta}
\def\epem{$e^+e^-$}
\def\KK{$K^0$-$\bar{K^0}$}
\def\wi{\omega_{\alpha}}
\def\xj{\chi_j}
\def\Wmu{W_\mu}
\def\Wnu{W_\nu}
\def\m#1{{\tilde m}_#1}
\def\mH{m_H}
\def\mw#1{{\tilde m}_{\omega #1}}
\def\mx#1{{\tilde m}_{\chi^{0}_#1}}
\def\mc#1{{\tilde m}_{\chi^{+}_#1}}
\def\mwi{{\tilde m}_{\omega i}}
\def\mxi{{\tilde m}_{\chi^{0}_i}}
\def\mci{{\tilde m}_{\chi^{+}_i}}
\def\mz{M_z}
\def\sw{\sin\theta_W}
\def\cw{\cos\theta_W}
\def\cb{\cos\beta}
\def\sb{\sin\beta}
\def\rwi{r_{\omega i}}
\def\rxj{r_{\chi j}}
\def\rfp{r_f'}
\def\Kik{K_{ik}}
\def\Fq2{F_{2}(q^2)}

%%%%%%%%%%%%%%%%%%%%%%%%%%%%%
%
\def\lsim{\ ^<\llap{$_\sim$}\ }
\def\gsim{\ ^>\llap{$_\sim$}\ }
\def\r2{\sqrt 2}
\def\beq{\begin{equation}}
\def\eeq{\end{equation}}
\def\beqn{\begin{eqnarray}}
\def\eeqn{\end{eqnarray}}
\def\rmuu{\gamma^{\mu}}
\def\rmud{\gamma_{\mu}}
\def\PL{{1-\gamma_5\over 2}}
\def\PR{{1+\gamma_5\over 2}}
\def\sinW2{\sin^2\theta_W}
\def\AEM{\alpha_{EM}}
\def\mul{M_{\tilde{u} L}^2}
\def\mur{M_{\tilde{u} R}^2}
\def\mdl{M_{\tilde{d} L}^2}
\def\mdr{M_{\tilde{d} R}^2}
\def\mz2{M_{z}^2}
\def\c2b{\cos 2\beta}
\def\au{A_u}         
\def\ad{A_d}
\def\cob{\cot \beta}
\def\v#1{v_#1}
\def\tb{\tan\beta}
\def\epem{$e^+e^-$}
\def\KK{$K^0$-$\bar{K^0}$}
\def\wi{\omega_{\alpha}}
\def\xj{\chi_j}
\def\Wmu{W_\mu}
\def\Wnu{W_\nu}
\def\m#1{{\tilde m}_#1}
\def\mH{m_H}
\def\mw#1{{\tilde m}_{\omega #1}}
\def\mx#1{{\tilde m}_{\chi^{0}_#1}}
\def\mc#1{{\tilde m}_{\chi^{+}_#1}}
\def\mwi{{\tilde m}_{\omega i}}
\def\mxi{{\tilde m}_{\chi^{0}_i}}
\def\mci{{\tilde m}_{\chi^{+}_i}}
\def\mz{M_z}
\def\sw{\sin\theta_W}
\def\cw{\cos\theta_W}
\def\cb{\cos\beta}
\def\sb{\sin\beta}
\def\rwi{r_{\omega i}}
\def\rxj{r_{\chi j}}
\def\rfp{r_f'}
\def\Kik{K_{ik}}
\def\Fq2{F_{2}(q^2)}
\def\f{\({\cal F}\)}
\def\d1{{\f(\tilde c;\tilde s;\tilde W)+ \f(\tilde c;\tilde \mu;\tilde W)}}
%%%%%%%%%%%%%%%%%%%%%%%%%%%%%%%%%%
\def\tw{\tan\theta_W}
\def\sec2w{sec^2\theta_W}
%%%%%%%%%%%%%%%%%%%%%%%%%%%%%%%%%%

\begin{titlepage}

\begin{center}
{\large {\bf Neutralino Exchange Corrections to the Higgs Boson  Mixings with Explicit CP Violation}}\\
\vskip 0.5 true cm
\vspace{2cm}
\renewcommand{\thefootnote}
{\fnsymbol{footnote}}
 Tarek Ibrahim$^{a}$ and Pran Nath$^{b}$  
\vskip 0.5 true cm
\end{center}

\noindent
{a. Department of  Physics, Faculty of Science,
University of Alexandria,}\\
{ Alexandria, Egypt}\\ 
{b. Department of Physics, Northeastern University,
Boston, MA 02115-5000, USA} \\
%\footnote{ $\dagger$ : Permanent address}
\vskip 1.0 true cm
\centerline{\bf Abstract}
\medskip
A calculus for the derivatives of the eigen values of the neutralino 
mass matrix with respect to the CP violating background fields is
developed and used to compute the mixings among the CP even and the
CP odd Higgs sectors arising from the inclusion of the neutralino sector 
consisting of the neutralino,  the Z boson, and the neutral 
Higgs bosons ($\chi^0_i-Z-h^0-H^0$) exchange in the loop  
contribution to the effective potential including the 
effects of large CP violating phases. Along with  the
stop, sbottom, stau and chargino-W-charged Higgs ($\chi^+-W-H^+$)
contributions computed previously the present analysis completes the 
 one loop corrections to the Higgs boson mass matrix
in the presence of large phases.
CP violation in the neutral Higgs sector is discussed in the above
framework with specific focus on the mixings of the CP even and the CP 
odd sectors arising from the neutralino sector.
It is shown that numerically the effects of the neutralino exchange 
contribution on the mixings of the CP even and the CP odd sectors are
comparable to the effects of the stop and of the chargino exchange 
contributions
and thus the neutralino exchange contribution must be included for
a realistic analysis of mixings in the CP even and the CP odd sectors. 
 Phenomenological implications of these results are discussed.
\end{titlepage}

\section{Introduction}
CP violation in supersymmetric theories via soft supersymmetry breaking
parameters has received a considerable degree of attention since
the beginning of the formulation of supersymmetric models\cite{ellis}.
Recently, there has been enhanced interest in the investigation of
their effects due to the realization that supersymmetric theories
may allow for large CP violating phases\cite{na} consistent with the 
electric dipole moment of the electron and of the neutron\cite{edmexp}. 
Such a situation can arise because of several possibilities,
such as the SUSY spectrum being heavy\cite{na}, due to internal 
cancellations\cite{in1}
and due to the possibility that the CP phases may reside in the
third generation and consequently their effects on the first 
two generation EMDs are suppressed\cite{chang}.
Of course it is possible that a more unified framework may determine
the combination of phases that enter the EDMs to be small\cite{bdm1}.
However, we shall investigate here the possibility that the phases
are large and the EDM constraints  are satisfied by one of the
methods discussed  above so that the sparticle spectrum is 
consistent with the naturalness constraints (see, e.g., Ref.\cite{ccn}).
 In this case their effects on low energy physics can be quite significant
and  a number of low energy phenomena have been discussed 
including the effect of CP phases. These include the effect of
CP phases on g-2\cite{ing2}, on dark matter\cite{olive},  on the
trileptonic signal\cite{trilep}, on baroyogenesis\cite{baryogenesis},
and on other low energy phenomena\cite{other}. 
Another area where the effect of CP phases has been discussed is 
the Higgs sector\cite{pilaftsis,pilaftsis2,demir,inhiggs,drees,carena}. 
It is well known that loop corrections to the
Higgs masses and mixings are important\cite{ellis0}. 
In fact in the absence of
the loop corrections the lightest Higgs boson mass must lie below 
$M_Z$ which is already exprimentally excluded and it is the presence
of the loop corrections that raises its value  above $M_Z$.
 An interesting phenomenon arises if the loop corrections have
 CP violating phases. In this case it has been pointed out that
 a significant mixing can occur between the CP even and the CP odd
 neutral Higgs sectors of the theory\cite{pilaftsis}. 
 In Refs.\cite{pilaftsis,pilaftsis2,demir,inhiggs,drees,carena} the effect
 of CP phases via the stop and sbottom exchanges was carried out.
 Further, in the work of Ref.\cite{inhiggs} it was pointed out that the
 effect of chargino loop corrections can be quite significant
 and in fact the CP effects from the chargino exchange may even 
 dominate the CP effects from the stop-sbottom exchange for 
 the case of large $\tan\beta$. 
 
  In this paper we give an analysis of the one loop
  correction to the Higgs boson mass including the neutralino-Z boson-neutral
  Higgs exchange including the CP violating phases. The inclusion 
  of the CP dependent neutralino exchange corrections are more 
  intricate relative to the stop-sbottom exchanges and the 
  chargino exchanges.
  This is due to the fact that the stop-sbottom exchange and 
   the chargino exchange involve diagonalization of only 
   $2\times 2$ squark and chargino mass matrices and thus the
   evaluation of their contribution can be carried out analytically
   in a straightforward fashion. For the case of the neutralino
   exchange the neutralino mass matrix is a $4\times 4$ object and
   its diagonalization analytically is more intricate and a 
   straightforward technique for the analysis is wieldy. In this
   paper we develop a calculus for the derivatives of the eigen values of
   the neutralino mass matrix to obtain an explicit analytic expression
   for the neutralino exchange contribution.
   The outline of the rest of the paper is as follows.
   In Sec.2 we give the Higgs potential and 
   discuss the minimization conditions in the presence of 
   the CP violating phases. In Sec.3 we discuss the calculus
   for the computation of derivatives of the eigen values of the 
   neutralino mass
   matrix. In Sec.4 we use the technique of Sec.3 and compute 
    the one loop contributions 
   to the Higgs boson mass matrix from the neutralino-Z-neutral Higgs
   boson exchange. Discussion of the numerical results is given
   in Sec.5. Conclusions are given in Sec.6. Some further details
   of the analysis are given in Appendices A and B.

\section{CP Phases and Minimization of Higgs Potential}
 We begin by defining the soft SUSY breaking parameters for the
 mSUGRA case\cite{applied}. Here the low energy physics
 for the CP conserving case is parametrized by $m_0$, $m_{\frac{1}{2}}$,
$A_0$, and $\tan\beta$ where $m_0$ is the universal
scalar mass,  $m_{\frac{1}{2}}$ is the universal  gaugino
mass, $A_0$ is the universal trilinear coupling, and 
$\tan\beta =\frac{v_2}{v_1}$ is the ratio of the Higgs VEVs,
where the VEV of $H_2$  gives mass to the up quarks and 
the VEV of $H_1$ gives mass to the down quarks and the leptons.
In the presence of CP violation
mSUGRA allows for only two CP violating phases which can be taken to be
$\theta_{\mu_0}$, and $\alpha_{A_0}$ where $\theta_{\mu_0}$
is the phase of the Higgs mixing parameter $\mu_0$ and $\alpha_{A_0}$ is the 
phase of $A_0$. The analysis of this paper, however, will be more general,
 valid for the MSSM parameter space.
The Higgs sector in MSSM at the one loop level is described by the 
scalar potential $V(H_1,H_2)=V_0+\Delta V$ where 
\beqn
V_0=m_1^2 |H_1|^2+m_2^2|H_2|^2 +(m_3^2 H_1.H_2 + H.C.)\nonumber\\
+\frac{(g_2^2+g_1^2)}{8}|H_1|^4+
\frac{(g_2^2+g_1^2)}{8}|H_2|^4
-\frac{g_2^2}{2}|H_1.H_2|^2
+\frac{(g_2^2-g_1^2)}{4}|H_1|^2|H_2|^2\nonumber\\ 
\Delta V=\frac{1}{64\pi^2}
 \sum_i c_{i} (2J_i+1)(-1)^{2J_i}(M^4_i(H_1,H_2)
 (log\frac{M^2_i(H_1,H_2)}{Q^2}-\frac{3}{2}))
\eeqn
Here $m_1^2=m_{H_1}^2+|\mu|^2, m_2^2=m_{H_2}^2+|\mu|^2,
 m_3^2=|\mu B|$
and $m_{H_{1,2}}$ and $B$ are the soft SUSY breaking parameters,
$\Delta V$ is the one loop correction to the effective 
potential\cite{coleman,arnowitt1} and includes contributions from
 all the fields 
that enter MSSM consisting of the standard model fields and their 
superpartners, i.e., the sfermions, the gauginos and higgsinos\cite{arnowitt1}. 
The sum over i in Eq.(1) runs over particles with spin $J_i$ 
and $c_{i}(2J_i+1)$ counts the degrees of the ith particle, 
and Q is the renormalization group running scale. 
It is well known that the one loop corrections to the effective potential
can make significant contributions to the Higgs vacuum expectation 
values in the minimization of the effective potential\cite{arnowitt1}.

 In general the effective potential depends on the CP violating phases
 and its minimization will lead to induced CP violating effects on the
 Higgs vacuum expectation values\cite{pilaftsis}. It is found convenient 
 to parameterize the Higgs VEVs in the presence of CP violating effects
 in the following form
\beqn
(H_1)= \left(\matrix{H_1^0\cr
 H_1^-}\right)
 =\frac{1}{\sqrt 2} 
\left(\matrix{v_1+\phi_1+i\psi_1\cr
             H_1^-}\right),~~
(H_2)= \left(\matrix{H_2^+\cr
             H_2^0}\right)
=\frac{e^{i\theta_H}}{\sqrt 2} \left(\matrix{H_2^+ \cr
             v_2+\phi_2+i\psi_2}\right)
\eeqn
where $\theta_H$ is in general non-vanishing as a consequence of the 
minimization conditions. 
Thus the minimization of the potential 
with respect to the fields $\phi_1, \psi_1,\phi_2, \psi_2$ 
gives 

\beqn
\frac{1}{v_2}(\frac{\partial \Delta V}{\partial \psi_1})_0=
m_3^2 \sin\theta_H\nonumber\\
-\frac{1}{v_1}(\frac{\partial \Delta V}{\partial \phi_1})_0=
m_1^2+\frac{g_2^2+g_1^2}{8}(v_1^2-v_2^2)+m_3^2 \tan\beta \cos\theta_H
\eeqn

and 

\beqn
\frac{1}{v_1}
(\frac{\partial \Delta V}{\partial \psi_2})_0=m_3^2 \sin\theta_H\nonumber\\
-\frac{1}{v_2}(\frac{\partial \Delta V}{\partial \phi_2})_0=
m_2^2-\frac{g_2^2+g_1^2}{8}(v_1^2-v_2^2)+m_3^2 cot\beta \cos\theta_H
\eeqn
In the above the subscript 0 stands for the fact that we 
are evaluating the relevant quantities 
at the point $\phi_1=\phi_2=\psi_1=\psi_2=0$.
We note in passing that in Eqs.(3) and (4) only one of the two equations 
that involve the variation with respect to $\psi_1$ and $\psi_2$ 
is independent\cite{demir}.

\section{Calculus for Derivatives of Eigen Values of Neutralino Mass
Matrix}
As mentioned in Sec.1, 
in previous analyzes computations of the CP dependent loop corrections
from the stop-sbottom and from the chargino- W- charged Higgs sectors
have been carried out. In these analyzes one was able to analytically 
obtain the eigen values by diagonalizing the $2\times 2$ squark matrices
and the $2\times 2$ charginio mass matrix and then differentiate them
analytically to obtain the loop correction to the Higgs mass matrix.
As also pointed out in Sec.1 
for the neutralino exchange case the situation is more difficult since
the neutralino mass matrix is a $4\times 4$ matrix and the analytic
solutions for the eigen values of the neutralino $(mass)^2$ matrix
are not easily obtained. 
Here we expand on a technique introduced in Ref.\cite{arnowitt1} to
derive a calculus for the derivatives of the eigen values for the
 neutralino mass matrix. This technique is valid for an 
arbitrary high order eigen value equation. We shall show that quite
remarkably even though one cannot analytically solve for the 
eigen  values one can analytically solve for the derivatives 
of the eigen values with respect to the background fields 
in terms of the eigen values and the parameters that appear in the
eigenvalue equation. To illustrate the 
procedure we consider an nth order eigen value equation

\beqn
F(\lambda)=Det(M^{\dagger}M -\lambda I)=
\lambda^n+c^{(n-1)}\lambda^{n-1}+c^{(n-2)}\lambda^{n-2}+..
+c^{(1)}\lambda  +c^{(0)}=0
\eeqn
Here the co-efficients are explicit functions of the background
fields  
\beq
\Phi_{\alpha}=\{\phi_1,\phi_2,\psi_1,\psi_2\} 
\eeq
while the eigen values are implicit functions of the
background fields
through the satisfaction of the eigen value equation.
Eq.(5) has n eigen values which we denote by $\lambda_{i}$
 ($i = 1,2,..,n$).
From  Eq.(5) it follows that
\beq
\frac{\partial\lambda_{i}}{\partial\Phi_{\alpha}}
=-(\frac{D_{\alpha}F}{D_{\lambda}F})_{\lambda=\lambda_{i}}
\eeq
and
\beq
\frac{\partial^2\lambda_i}{\partial\Phi_{\alpha}\partial\Phi_{\beta}}
=[-\frac{D_{\alpha}FD_{\beta}FD^2_{\lambda}F}{(D_{\lambda}F)^3}
+\frac{D_{\alpha}FD_{\beta}D_{\lambda}F+D_{\beta}FD_{\alpha}
D_{\lambda}F}{(D_{\lambda}F)^2}
-\frac{D_{\alpha}D_{\beta}F}{D_{\lambda}F}]_{\lambda=\lambda_{i}}
\eeq
where $D_{\lambda}$ differentiates the $\lambda$ dependence in
$F$
\beq
D_{\lambda}F(\lambda)=\frac{dF}{d\lambda}
\eeq
and $D_{\alpha}$ differentiates only the co-efficients in Eq.(5),
i.e., 
\beq
D_{\alpha}F=c^{(n-1)}_{\alpha}\lambda^{(n-1)}+
c^{(n-2)}_{\alpha}\lambda^{(n-2)}+..+c^{(1)}_{\alpha}\lambda 
+c^{(0)}_{\alpha}
\eeq
$D_{\alpha}D_{\beta} F$ are similarly defined where
 $c_{\alpha}^{(k)}$ etc are replaced with 
$c_{\alpha\beta}^{(k)}$ where
\beq
c_{\alpha}^{(k)}=\frac{\partial c^{(k)}}{\partial\Phi_{\alpha}},~~~
c_{\alpha\beta}^{(k)}=\frac{\partial^2 c^{(k)}}
{\partial\Phi_{\alpha}\partial\Phi_{\beta}}
\eeq
and  the derivatives $D_{\alpha}D_{\lambda}$ are 
defined in an obvious way. We note in passing that $D_{\alpha}$ 
and $D_{\lambda}$ commute
\beq
[D_{\alpha},D_{\lambda}]=0
\eeq
Eqs.(7) and (8) are the central equations of our analysis.
It is easy to check that for the $2\times 2$ matrix case,
e.g., for the stop and the chargino exhanges, they give exactly
the results gotten by explicit differentiation of the eigen values.
However, now these equations provide us with a technique of 
analyzing cases where the analytic solutions to the eigen values
are not available.

\section{Neutralino, Z and neutral Higgs loop contributions}
As mentioned above the CP dependent contributions to the Higgs boson
masses  from stop and sbottom exchanges  have been discussed at length
in the literature\cite{pilaftsis,pilaftsis2,demir,inhiggs,drees,carena}. More  
recently the CP dependent chargino-W-charged Higgs contributions were
 also discussed\cite{inhiggs}. 
 In this work we use the technique discussed in Sec.3 
 to compute the contribution from the neutralino -Z - neutral Higgs
 exchange. The loop correction in this sub sector is given by  

\beqn
\Delta V(\chi^0_i,Z, h^0, H^0)=\frac{1}{64\pi^2}
(\sum_{i=1}^{4}(-2)M_{\chi_i^0}^4(log\frac{M_{\chi_i^0}^2}{Q^2}-\frac{3}{2})
+3 M_Z^4 (log\frac{M_Z^2}{Q^2}-\frac{3}{2})\nonumber\\
+  M_{h^0}^4 log(\frac{M_{h^0}^2}{Q^2}-\frac{3}{2})
+  M_{H^0}^4 log(\frac{M_{H^0}^2}{Q^2}-\frac{3}{2}))
\eeqn
The neutralino mass matrix is given by 
\beq
M_{\chi^0}=
\left(\matrix{\tilde m_1 & 0 & -\frac{g_1}{\sqrt 2} H_1^0 &
\frac{g_1}{\sqrt 2} H_2^0 \cr
0 & \tilde m_2 & \frac{g_2}{\sqrt 2} H_1^0 & -\frac{g_2}{\sqrt 2} H_2^0\cr
-\frac{g_1}{\sqrt 2} H_1^0 & \frac{g_2}{\sqrt 2} H_1^0  & 0 & -\mu \cr
\frac{g_1}{\sqrt 2} H_2^0 & -\frac{g_2}{\sqrt 2} H_2^0 & -\mu &0}\right)
\eeq
where $\mu=|\mu|e^{i\theta_{\mu}}$,  $\tilde m_1=|\tilde m_1|e^{i\xi_1}$
and $\tilde m_2=|\tilde m_2|e^{i\xi_2}$.
We note that in the supersymmetric limit $M_{\chi_{i}^0}$
$= (0,0, M_Z,M_Z)$ and $(M_{h^0}, M_{H^0})$=$(M_Z,0)$
and consequently in this limit the loop corrections from this sub sector
vanish. We return now to the full analysis and follow the method
described in Ref.\cite{inhiggs} to minimize the potential and 
compute the loop corrections. First we give the determination of 
$\theta_H$ from the minimization constraints including the stop,
the sbottom, the stau, the chargino and neutralino  contributions.
One finds that $\theta_H$ is given by
the equation

\beqn
m_3^2 \sin\theta_H =\frac{1}{2} \beta_{h_t} |\mu| |A_t| \sin\gamma_t
f_1(m_{\tilde t_1}^2, m_{\tilde t_2}^2)
+\frac{1}{2}  \beta_{h_b}|\mu| |A_b| \sin\gamma_b
f_1(m_{\tilde b_1}^2, m_{\tilde b_2}^2)\nonumber\\
+\frac{1}{2}  \beta_{h_{\tau}}|\mu| |A_{\tau}| \sin\gamma_{\tau}
f_1(m_{\tilde {\tau}_1}^2, m_{\tilde {\tau}_2}^2)
-\frac{g^2_2}{16\pi^2}|\mu| |\tilde m_2| \sin\gamma_2
f_1(m_{\tilde \chi_1}^2, m_{\tilde \chi_2}^2)\nonumber\\
+\frac{1}{16\pi^2} \sum_{j=1}^{4}\frac{M_{\chi_j^0}^2}{D_j}
(ln(\frac{M_{\chi_j^0}^2}{Q^2})-1)
 (M_{\chi_j^0}^4(-g_2^2|\mu| |\tilde m_2|\sin\gamma_2   
-g_1^2|\mu||\tilde m_1|\sin\gamma_1)\nonumber\\
+ M_{\chi_j^0}^2(g_2^2(|\tilde m_1|^2+|\mu|^2)|\tilde m_2|
|\mu|\sin\gamma_2
+g_1^2 (|\tilde m_1|^2+|\mu|^2)|\tilde m_1||\mu|\sin\gamma_1)\nonumber\\
+(-g_2^2 |\tilde m_1|^2 |\mu|^3 |\tilde m_2| \sin\gamma_2
-g_1^2 |\tilde m_2|^2 |\mu|^3 |\tilde m_1| \sin\gamma_1)) 
\eeqn
where 
\beqn
D_j\equiv (D_{\lambda}F)_{\lambda = \lambda_j}
 =4M_{\chi_j^0}^6+3a M_{\chi_j^0}^4+2b M_{\chi_j^0}^2 +c\nonumber\\
 \beta_{h_t}=\frac{3 h_t^2}{16\pi^2},
~~ \beta_{h_b}=\frac{3 h_b^2}{16\pi^2},
\beta_{h_{\tau}}= \frac{3 h_{\tau}^2}{16\pi^2}\nonumber\\
\gamma_t=\alpha_{A_t} + \theta_{\mu}, ~~
\gamma_b=\alpha_{A_b} + \theta_{\mu},~~
\gamma_{\tau}=\alpha_{A_{\tau}}+\theta_{\mu},~~ 
\gamma_1=\xi_1 + \theta_{\mu}, \gamma_2=\xi_2 + \theta_{\mu}
\eeqn
and where a,b,c are defined in Appendix A 
and $f_1(u,v)$ is given by
\beq
f_1(u,v)=-2+log\frac{uv}{Q^4} + \frac{v+u}{v-u}log\frac{v}{u}
\eeq
To construct the mass squared matrix of the Higgs scalars 
we need to compute the quantity
\beq
M_{\alpha\beta}^2=(\frac{\partial^2 V}{\partial \Phi_{\alpha}\partial\Phi_{\beta}})_0
=M_{\alpha\beta}^{2(0)}+ \Delta M_{\alpha\beta}^2
\eeq
where $M_{\alpha\beta}^{2(0)}$ is the contribution from $V_0$ and 
$\Delta M_{\alpha\beta}^2$ is the contribution from $\Delta V$ 
where $\Phi_{\alpha} (\alpha=1-4)$ are defined by Eq.(6)
and as already mentioned earlier the subscript 0  means that we set 
$\phi_1=\phi_2=\psi_1=\psi_2=0$ after evaluating the mass matrix.
The loop contribution $\Delta M_{\alpha\beta}^2$ arising from the neutralino-
Z-neutral Higgs sector is given by 
 
\beq
\Delta M_{\alpha\beta}^2=
\frac{1}{32\pi^2}
Str(\frac{\partial M^2}{\partial \Phi_{\alpha}}\frac{\partial M^2}{\partial\Phi_{\beta}}
log\frac{M^2}{Q^2}+M^2 \frac{\partial^2 M^2}{\partial \Phi_{\alpha}\partial \Phi_{\beta}}
(log\frac{M^2}{Q^2}-1))_0
\eeq
Computation of the $4\times 4$ Higgs  mass matrix in the basis 
of Eq.(6) gives

\beq
\left(\matrix{M_Z^2c_{\beta}^2+M_A^2s_{\beta}^2+\Delta_{11} &
-(M_Z^2+M_A^2)s_{\beta}c_{\beta}+\Delta_{12} &\Delta_{13}s_{\beta}&\Delta_{13}
 c_{\beta}\cr
-(M_Z^2+M_A^2)s_{\beta}c_{\beta}+\Delta_{12} &
M_Z^2s_{\beta}^2+M_A^2c_{\beta}^2+\Delta_{22} & \Delta_{23} s_{\beta}
&\Delta_{23} c_{\beta}\cr
\Delta_{13} s_{\beta} & \Delta_{23} s_{\beta}&(M_A^2+\Delta_{33})s_{\beta}^2 & 
(M_A^2+\Delta_{33})s_{\beta}c_{\beta}\cr
\Delta_{13} c_{\beta} &\Delta_{23} c_{\beta} &(M_A^2+\Delta_{33})s_{\beta}c_{\beta} & 
(M_A^2+\Delta_{33})c_{\beta}^2}\right)
\eeq
where $c_{\beta}(s_{\beta})=\cos\beta(\sin\beta)$ and $m_A^2$
is given by
\beqn
m_A^2=(\sin\beta\cos\beta)^{-1}(-m_3^2\cos\theta +\frac{1}{2}\beta_{h_t} 
|A_t||\mu|\cos\gamma_t f_1(m_{\tilde t_1}^2,m_{\tilde t_2}^2)\nonumber\\
+\frac{1}{2}\beta_{h_b} |A_b||\mu| \cos\gamma_b f_1
(m_{\tilde b_1}^2,m_{\tilde b_2}^2) +\frac{1}{2}\beta_{h_{\tau}} |A_{\tau}||\mu| \cos\gamma_{\tau} f_1
(m_{\tilde {\tau}_1}^2,m_{\tilde {\tau}_2}^2) \nonumber\\
+\frac{g_2^2}{16\pi^2}|\tilde m_2|
|\mu| \cos\gamma_2 f_1(m_{\chi_1^+}^2, m_{\chi_2^+}^2)\nonumber\\
-\frac{1}{16\pi^2}\sum_{j=1}^{4}\frac{M_{\chi_j}^2}{D_j} 
(log(\frac{M_{\chi_j}^2}{Q^2}) -1)
[M_{\chi_j}^4(-g_2^2|\mu||\tilde m_2|\cos\gamma_2 -g_1^2|\mu||\tilde m_1|
\cos\gamma_1)\nonumber\\
+M_{\chi_j}^2(g_2^2(|\tilde m_1|^2+|\mu|^2)|\mu| |\tilde m_2|\cos\gamma_2
+g_1^2(|\tilde m_2|^2+|\mu|^2)|\mu| |\tilde m_1|\cos\gamma_1)\nonumber\\
-g_2^2 |\tilde m_1|^2 |\mu|^3 |\tilde m_2| \cos\gamma_2
-g_1^2 |\tilde m_2|^2 |\mu|^3 |\tilde m_1| \cos\gamma_1] )
\eeqn  
The first term in the second 
 brace on the right hand side of Eq.(21) is the tree term, 
while the second, the
third, the fourth and the fifth terms come from the stop, sbottom, 
stau and chargino
exchange contributions. The remaining contributions in Eq.(21) arise
from the neutralino sector.  The $\Delta$'s appearing in Eq.(20)
can be decomposed as follows
\beq
\Delta_{\alpha\beta}=\Delta_{\alpha\beta\tilde t}+\Delta_{\alpha\beta\tilde b}+\Delta_{\alpha\beta\tilde {\tau}}+\Delta_{\alpha\beta\chi^+} + \Delta_{\alpha\beta\chi^0}
\eeq
where $\Delta_{\alpha\beta\tilde t}$ is the contribution from the 
stop (and top) exchange in the loops, $\Delta_{\alpha\beta\tilde b}$ is the 
contribution from the sbottom (and bottom) exchange  in the loops,
$\Delta_{\alpha\beta\tilde {\tau}}$ is the contribution from 
the stau (and tau) exchange, 
$\Delta_{\alpha\beta\chi^+}$ is the contribution from the chargino 
 (and W and charged Higgs) exchange in the loops,
   and $\Delta_{\alpha\beta\chi^0}$ is the contribution
 arising from the neutralino (and Z and neutral Higgs exchange)
 in the loops.   The computations of
$\Delta_{\alpha\beta\tilde t}$, $\Delta_{\alpha\beta\tilde b}$, 
$\Delta_{\alpha\beta \tilde \tau}$,
and $\Delta_{\alpha\beta\chi^+}$
have been given before and would  not  be reproduced  here.
We compute here only the $ \Delta_{\alpha\beta\chi^0}$  arising from
the ($\chi^0_i-Z-h^0-H^0$) exchange. The $ \Delta_{\alpha\beta\chi^0}$
 are listed below.

\beqn
\Delta_{11\chi^0}=-\frac{1}{16\pi^2}\sum_{j=1}^{4} M_{\chi_j}^2
(ln(\frac{M_{\chi_j}^2}{Q^2})-1)\nonumber\\
\{-\frac{(a_1 M_{\chi_j}^6+b_1M_{\chi_j}^4+c_1M_{\chi_j}^2+d_1)^2
(12M_{\chi_j}^4+6aM_{\chi_j}^2+2b)}{D_j^3}\nonumber\\
+ \frac{2(a_1 M_{\chi_j}^6+b_1M_{\chi_j}^4+c_1M_{\chi_j}^2+d_1)
(3a_1 M_{\chi_j}^4+2b_1M_{\chi_j}^2+c_1)}{D_j^2}\}\nonumber\\
-\frac{1}{16\pi^2}\sum_{j=1}^{4} \frac{(a_1 M_{\chi_j}^6+
b_1M_{\chi_j}^4+c_1
M_{\chi_j}^2+d_1)^2}{D_j^2}ln(\frac{M_{\chi_j}^2}{Q^2})
+\frac{3}{128\pi^2}(g_1^2+g_2^2)^2 v_1^2 ln(\frac{M_Z^2}{Q^2})\nonumber\\
-\frac{1}{32\pi^2}(\frac{1}{16}\frac{A_0^2}{(M_{H^0}^2-M_{h^0}^2)^2}
f_2(M_{H^0}^2, M_{h^0}^2)
-\frac{1}{16}(g_1^2+g_2^2)^2 v_1^2ln\frac{M_{H^0}^2M_{h^0}^2}{Q^4}\nonumber\\
-\frac{1}{8}(g_1^2+g_2^2) \frac{v_1 A_0}{(M_{H^0}^2-M_{h^0}^2)}
ln\frac{M_{H^0}^2}{M_{h^0}^2})
\eeqn

where $D_j$ is defined in Eq.(16) and $f_2$ is defined by 
\beq 
f_2(u,v)=-2+\frac{v+u}{v-u}ln\frac{v}{u}
\eeq

\beqn
\Delta_{22\chi^0}=-\frac{1}{16\pi^2}\sum_{j=1}^{4} M_{\chi_j}^2
(ln(\frac{M_{\chi_j}^2}{Q^2})-1)\nonumber\\
\{ -\frac{(a_2M_{\chi_j}^6+ b_2M_{\chi_j}^4+c_2M_{\chi_j}^2+d_2)^2
(12M_{\chi_j}^4+6aM_{\chi_j}^2+2b)}{D_j^3}\nonumber\\
+ \frac{2(a_2M_{\chi_j}^6+b_2M_{\chi_j}^4+c_2M_{\chi_j}^2+d_2)
(3a_2M_{\chi_j}^4+2b_2M_{\chi_j}^2+c_2)}{D_j^2}\}\nonumber\\
-\frac{1}{16\pi^2}\sum_{j=1}^{4} \frac{(a_2 M_{\chi_j}^6+
b_2M_{\chi_j}^4+c_2
M_{\chi_j}^2+d_2)^2}{D_j^2}ln(\frac{M_{\chi_j}^2}{Q^2})
+\frac{3}{128\pi^2}(g_1^2+g_2^2)^2 v_2^2 ln(\frac{M_Z^2}{Q^2})\nonumber\\
-\frac{1}{32\pi^2}(\frac{1}{16}\frac{B_0^2}{(M_{H^0}^2-M_{h^0}^2)^2}
f_2(M_{H^0}^2, M_{h^0}^2)
-\frac{1}{16}(g_1^2+g_2^2)^2 v_2^2ln\frac{M_{H^0}^2M_{h^0}^2}{Q^4}\nonumber\\
-\frac{1}{8}(g_1^2+g_2^2) \frac{v_2B_0}{(M_{H^0}^2-M_{h^0}^2)}
ln\frac{M_{H^0}^2}{M_{h^0}^2})
\eeqn

\beqn
\Delta_{12\chi^0}=-\frac{1}{16\pi^2}\sum_{j=1}^{4} M_{\chi_j}^2
(ln(\frac{M_{\chi_j}^2}{Q^2})-1)\nonumber\\
\{-\frac{(a_1 M_{\chi_j}^6+b_1M_{\chi_j}^4+c_1M_{\chi_j}^2+d_1)
(a_2 M_{\chi_j}^6+b_2M_{\chi_j}^4+c_2M_{\chi_j}^2+d_2)
(12M_{\chi_j}^4+6aM_{\chi_j}^2+2b)}{D_j^3}\nonumber\\
+ \frac{(a_1 M_{\chi_j}^6+b_1M_{\chi_j}^4+c_1M_{\chi_j}^2+d_1)
(3a_2 M_{\chi_j}^4+2b_2M_{\chi_j}^2+c_2)}{D_j^2}\nonumber\\
+\frac{
(a_2 M_{\chi_j}^6+b_2M_{\chi_j}^4+c_2M_{\chi_j}^2+d_2)
(3a_1 M_{\chi_j}^4+2b_1M_{\chi_j}^2+c_1)}{D_j^2}\}\nonumber\\
-\frac{1}{16\pi^2}\sum_{j=1}^{4} \frac{(a_1 M_{\chi_j}^6+b_1M_{\chi_j}^4+c_1M_{\chi_j}^2+d_1)
(a_2 M_{\chi_j}^6+b_2M_{\chi_j}^4+c_2M_{\chi_j}^2+d_2)}{D_j^2}
ln(\frac{M_{\chi_j}^2}{Q^2})\nonumber\\
+\frac{3}{128\pi^2}(g_1^2+g_2^2)^2 v_1v_2 ln(\frac{M_Z^2}{Q^2})\nonumber\\
-\frac{1}{32\pi^2}(\frac{1}{16}\frac{A_0B_0}{(M_{H^0}^2-M_{h^0}^2)^2}
f_2(M_{H^0}^2, M_{h^0}^2)
-\frac{1}{16}(g_1^2+g_2^2)^2 v_1v_2ln\frac{M_{H^0}^2M_{h^0}^2}{Q^4}\nonumber\\
-\frac{1}{16}(g_1^2+g_2^2) \frac{v_1B_0+v_2A_0}{(M_{H^0}^2-M_{h^0}^2)}
ln\frac{M_{H^0}^2}{M_{h^0}^2})
\eeqn

\beqn
\Delta_{13\chi^0}=-\frac{1}{16\pi^2}\sum_{j=1}^{4} M_{\chi_j}^2
(ln(\frac{M_{\chi_j}^2}{Q^2})-1)\frac{1}{\sin\beta}\nonumber\\
\{-\frac{(a_1 M_{\chi_j}^6+b_1M_{\chi_j}^4+c_1M_{\chi_j}^2+d_1)
(a_3 M_{\chi_j}^6+b_3M_{\chi_j}^4+c_3M_{\chi_j}^2+d_3)
(12M_{\chi_j}^4+6aM_{\chi_j}^2+2b)}{D_j^3}\nonumber\\
+ \frac{(a_1 M_{\chi_j}^6+b_1M_{\chi_j}^4+c_1M_{\chi_j}^2+d_1)
(3a_3 M_{\chi_j}^4
+2b_3M_{\chi_j}^2+c_3)}{D_j^2}\nonumber\\
+\frac{(a_3 M_{\chi_j}^6+b_3M_{\chi_j}^4+c_3M_{\chi_j}^2+d_3)
(3a_1 M_{\chi_j}^4+2b_1M_{\chi_j}^2+c_1)}{D_j^2}\}\nonumber\\
-\frac{1}{16\pi^2}\sum_{j=1}^{4}\frac{1}{\sin\beta}
 \frac{(a_1 M_{\chi_j}^6+b_1M_{\chi_j}^4+c_1M_{\chi_j}^2+d_1)
(a_3 M_{\chi_j}^6+b_3M_{\chi_j}^4+c_3M_{\chi_j}^2+d_3)}{D_j^2}
ln(\frac{M_{\chi_j}^2}{Q^2})
\eeqn

\beqn
\Delta_{23\chi^0}=-\frac{1}{16\pi^2}\sum_{j=1}^{4} M_{\chi_j}^2
(ln(\frac{M_{\chi_j}^2}{Q^2})-1)\frac{1}{\cos\beta}\nonumber\\
(-\frac{(a_3'M_{\chi_j}^6+b_3'M_{\chi_j}^4+c_3'M_{\chi_j}^2+d_3')
(a_2M_{\chi_j}^6+b_2M_{\chi_j}^4+c_2M_{\chi_j}^2+d_2)
(12M_{\chi_j}^4+6aM_{\chi_j}^2+2b)}{D_j^3}\nonumber\\
+ \frac{(a_3'M_{\chi_j}^6+b_3'M_{\chi_j}^4+c_3'M_{\chi_j}^2+d_3')
(3a_2M_{\chi_j}^4+2b_2M_{\chi_j}^2+c_2)}{D_j^2}\nonumber\\
+\frac{
(a_2M_{\chi_j}^6+b_2M_{\chi_j}^4+c_2M_{\chi_j}^2+d_2)
(3a_3'M_{\chi_j}^4+2b_3'M_{\chi_j}^2+c_3')}{D_j^2}\nonumber\\
-\frac{1}{16\pi^2}\sum_{j=1}^{4}\frac{1}{\cos\beta}
 \frac{(a_3'M_{\chi_j}^6+b_3'M_{\chi_j}^4+c_3'M_{\chi_j}^2+d_3')
(a_2M_{\chi_j}^6+b_2M_{\chi_j}^4+c_2M_{\chi_j}^2+d_2)}{D_j^2}
ln(\frac{M_{\chi_j}^2}{Q^2})
\eeqn

\beqn
\Delta_{33\chi^0}=-\frac{1}{16\pi^2}\sum_{j=1}^{4} M_{\chi_j}^2
(ln(\frac{M_{\chi_j}^2}{Q^2})-1)\nonumber\\
(-\frac{(a_3'M_{\chi_j}^6+b_3'M_{\chi_j}^4+c_3'M_{\chi_j}^2+d_3')^2
(12M_{\chi_j}^4+6aM_{\chi_j}^2+2b)}{D_j^3}\frac{1}{\cos^2\beta}\nonumber\\
+ \frac{2(a_3'M_{\chi_j}^6+b_3'M_{\chi_j}^4+c_3'M_{\chi_j}^2+d_3')
(3a_3'M_{\chi_j}^4+2b_3'M_{\chi_j}^2+c_3')}{D_j^2}\frac{1}{\cos^2\beta}\nonumber\\
-\frac{1}{16\pi^2}\sum_{j=1}^{4} \frac{1}{\cos^2\beta}
\frac{(a_3'M_{\chi_j}^6+
b_3'M_{\chi_j}^4+c_3'M_{\chi_j}^2+d_3')^2}{D_j^2}
ln(\frac{M_{\chi_j}^2}{Q^2})
\eeqn
The parameters a,b,c and the derivatives $a_i,b_i,c_i,d_i$
(i=1,2, etc.) that appear in Eqs.(23-29) are defined in Appendices
A and B. 
Eqs.(23-29) constitute the main new theoretical results of this
paper. These results along with the computations of 
$\Delta_{\alpha\beta\tilde t}$, $\Delta_{\alpha\beta\tilde b}$, 
$\Delta_{\alpha\beta\tilde {\tau}}$
and 
$\Delta_{\alpha\beta\chi^+}$ give a complete determination of the  
CP dependent one loop  contributions to the Higgs boson masses and
mixings. As has been noted before  it is preferable to 
work with a $3\times 3$ matrix rather than the $4\times 4$ matrix of
Eq.(20). The desired $3\times 3$ matrix can be gotten from Eq.(20) 
by going to the basis  
\beqn
\psi_{1D}=\sin\beta \psi_1+ \cos\beta \psi_2, ~~~
\psi_{2D}=-\cos\beta \psi_1+\sin\beta \psi_2
\eeqn
In this basis the field $\psi_{2D}$ is the zero mass Goldstone boson
and decouples while the remaining $(mass)^2$ matrix in the basis 
$\phi_1, \phi_2, \psi_{1D}$ is given by 
\beq
M^2_{Higgs}=
\left(\matrix{M_Z^2c_{\beta}^2+M_A^2s_{\beta}^2+\Delta_{11} &
-(M_Z^2+M_A^2)s_{\beta}c_{\beta}+\Delta_{12} &\Delta_{13}\cr
-(M_Z^2+M_A^2)s_{\beta}c_{\beta}+\Delta_{12} &
M_Z^2s_{\beta}^2+M_A^2c_{\beta}^2+\Delta_{22} & \Delta_{23} \cr
\Delta_{13} & \Delta_{23} &(M_A^2+\Delta_{33})}\right)
\eeq
We label the eigen values for this case $m_{H_1}^2$, $m_{H_2}^2$,
$m_{H_3}^2$ corresponding to the eigen states $H_1, H_2, H_3$.
These eigen states are in general admixtures of the CP even and the
CP odd states
due to  the mixing generated by $\Delta_{13}$ and $\Delta_{23}$.
Thus the CP even-odd mixings arise from $\Delta_{13}$ and $\Delta_{23}$
and these are  nonvanishing only in the presence of CP violation and 
vanish when the phases go to zero and one recovers the usual result of two 
distinct  (one CP even and the other CP odd) Higgs sectors.
We note in passing that $\Delta_{33}$ also vanishes in the limit 
when the CP phases go to zero. This was also the behavior that was observed
when the contributions from the stop, sbottom, stau and chargino exchanges
were considered.  Since the main point of this work
is to study the phenomenon of CP even-odd mixing the main focus of 
our analysis is the computation of $\Delta{ij}$ and specifically
of $\Delta_{13}$ and $\Delta_{23}$ which are the basic sources of 
mixings between the CP even and the CP odd sectors.
We order the eigen values of Eq.(31) in such a way that in the limit of no
CP violation one has 
$(m_{H_1}, m_{H_2},m_{H_3})$$\rightarrow$ 
$(m_H, m_h, m_A)$ and $(H_1,H_2,H_3)$$\rightarrow$ $(H, h, A)$
where (h, H)   are (light, heavy) CP even Higgs and A is the CP 
odd Higgs in the absence of CP violation.

\section{Discussion of the Neutralino Exchange Contribution to
CP even CP odd Higgs Mixing}
The analytical results given above are quite general as they
apply to the MSSM parameter space. However, the MSSM parameter
space is quite large. Thus for a numerical study of the CP
effects including those from the neutralino sector we will work
with a constrained set of parameters consisting of the
parameter space $m_0, m_{\frac{1}{2}}$, $m_A$, $|A_0|$, $\tan\beta$,
$\theta_{\mu}$, $\alpha_{A_0}$, $\xi_1$, $\xi_2$ and $\xi_3$.
Starting with these all other low energy parameters are obtained
by a renormalization group evolution by running the parameters
from the GUT scale down to the electro-weak scale. Of course
one is free to utilize the formulae derived above for the
more general MSSM parameter space. As discussed in Sec.1 one can
satisfy the EDM constraints in the presence of large phases.
This can come about in a variety of ways. As pointed  out in Sec.1  
one possibility is that
the internal cancellations can occur which allow for large phases
consistent with the EDM constraints. The other possibility is that
that CP phases appear only in the third generation which
suppresses their contributions to the EDMs of the quarks and the
leptons in the first two generations to achieve consistency with
the experimental constraints. There also exist scenarios which are
linear combinations of these two. For the purpose of this analysis
we do not revisit the problem of the satisfaction of the EDM constraints.
Rather we shall assume that regions of the parameter space exist
where such constraints are satisfied and examine the effect of the
phases on the Higgs masses and mixing. Specifically we are interested
in the effects of the neutralino exchange contributions on 
$\Delta_{13}$ and $\Delta_{23}$, and thus, on the mixings of the
CP even and the CP sectors. 

It was pointed out in Sec.4 that the neutralino, the Z and the neutral Higgs
exchanges together form a sub sector so that in the supersymmetric limit
one finds that the one loop correction to the effective potential 
from this sub sector vanishes. This phenomenon is similar to what was
also seen in the exchange of the chargino, the W and the charged Higgs
where the contribution from that sector to the effective potential 
vanishes in the supersymmetric limit. It was also seen in the analysis
of the chargino-W-charged Higgs exchange that the CP even-odd mixing
arising from this sector was roughly Q independent because of the 
sum of the three separate contributions within this sector.
A very similar situation is also realized in the neutralino sector.
Here again because of the contributions from the neutralino, the Z and
the neutral Higgs exchanges their sum contribution to the CP even-odd 
mixing is roughly scale independent. However, unlike the chargino-W-
charged Higgs exchange where one could demonstrate the above phenomenon
analytically, here one has to demonstrate it numerically due to the 
more analytically complex nature of the results.This is exhibited in
Fig.1 where a plot the percentage of the CP even
     component $\phi_1$ and the CP odd component $\psi_{1D}$ of $H_1$ as 
     a function of Q is given. The analysis shows an approximate 
     independence in Q of the CP even-odd mixing. We turn now to a 
     discussion of other aspects of the analysis below.

In Fig.2  we plot the quantity  $\Delta_{13}$
as a function of the CP phase of the U(1) gaugino mass $\xi_1$. 
The  plots exhibited  in Fig.2 contain 
the stop, the sbottom, the stau the chargino and the neutralino exchange contributions.
Among  the above exchanges 
the neutralino exchange contribution is the only one that
depends on $\xi_1$, and thus the variation of $\Delta_{13}$ with $\xi_1$ 
arises only from this exchange. From Fig.2 the size of the neutralino 
exchange contribution can be seen to be fairly substantial. Specifically,
the analysis of Fig.2 shows that the neutralino exchange contribution is
comparable to the effects from the stop and chargino exchanges. 
A plot of $\Delta_{23}$ vs $\xi_1$ is given Fig.3. As in Fig.2 one finds
that $\Delta_{23}$ is quite sensitive to the CP violating phase $\xi_1$.
As in Fig.2 here again the neutralino exchange contribution is 
comparable to the stop and the chargino exchange contribution.
  An analysis of the percentage of the
 CP even component $\phi_1$ of $H_1$ (upper curves) and of the percentage
 of the CP odd component $\psi_{1D}$ of $H_1$ (lower curves) arising from
 the exchange of the stop, the sbottom, the stau, the chargino and the neutralino 
 sector contributions  as a function of $\xi_1$  is given in Figs.4.
 As expected from the analysis of Fig.2 and Fig.3 one finds that 
 there is a significant mixing between the CP even and the CP odd 
 components of $H_1$. Further, as also expected from the analysis 
 of Figs 2 and 3, the CP even and CP odd components of $H_1$ 
 show a reasonably strong dependence on $\xi_1$. 

An analysis of the CP even and CP odd mixing in $H_1$ as a function of 
the SU(2) gaugino phase is given in Fig.5. Unlike Figs.2-4, where the
entire $\xi_1$ dependence arose from the neutralino exchange contribution
here the $\xi_2$ dependence of the CP even and CP odd components of 
$H_1$ arises from two sources, i.e., from the chargino and the neutralino
exchange contributions. Because of this the dependence of the CP even 
and CP odd components on $\xi_2$ is much stronger than on $\xi_1$ as
may be seen by comparing the plots of Figs. 2-4 with the plots of Fig.5.   
 In Fig.6 a plot of the percentage of the
 CP even component $\phi_1$ of $H_1$ (upper sets) and the CP odd 
 component $\psi_{1D}$ of $H_1$ (lower sets) arising from the
 exchange of the stop, the sbottom, the stau, the chargino and the neutralino 
 sector contributions is given as a function of $\theta_{\mu}$.
 In this case we find that the dependence of the CP even and the CP odd
 components on $\theta_{\mu}$ is also very strong. Indeed in this case 
 the mixings between the CP even and the CP odd states can be maximal 
 depending on the value of $\theta_{\mu}$. The strong dependence
 on $\theta_{\mu}$ can be understood as due to the fact that all 
 contributions, i.e., the stop, the sbottom, the stau, the chargino,
  and the neutralino
 contributions, depend on $\theta_{\mu}$. This in contrast to the
 dependence on $\xi_1$ which arises only from the neutralino exchange.

 Finally, in Fig.7 we give an analysis  of the percentage of the
 CP even component $\phi_1$ of $H_1$ (upper sets) and the CP odd 
 component $\psi_{1D}$ of $H_1$ (lower sets) arising from the
 exchange of the stop, the sbottom, the stau, the chargino and the neutralino 
 sector contributions as a function of $\tan\beta$. We find that the
 CP even and the CP odd mixings show a strong dependence on $\tan\beta$.
A similar strong dependence on $\tan\beta$ was seen also 
 in previous analyses\cite{inhiggs}. 
We note that the inclusion  of the neutralino contribution 
further sharpens the $\tan\beta$ dependence and one finds that the 
CP even (odd) component can vary from 100\% (0\%) to less than 60\%
(more than 40\%) as $\tan\beta$ is varied. 
 This sharper behavior of the amplitudes with $\tan\beta$ arises from the
 additional contributions from the neutralino, the neutral Higgs and 
 the $Z$ boson exchanges.
 An analysis similar to the above can be
 carried out for the case of the $H_2$ and $H_3$ fields. In the
 analysis of chargino exchange contributions it was  found that
 the CP odd component of $H_2$ is rather small while the analysis
 of $H_3$ parallels the analysis of $H_1$ with the only difference
 that  the roles of the CP even and the CP odd components is reversed.
 Much the same situations occurs in this case and thus we omit the detailed
 discussion of these states. 
  
\section{Conclusions}
In this paper we have developed a calculus for the derivatives of 
the eigen values of the neutralino mass matrix with respect to the
background fields which are in general dependent on CP violating 
phases. The calculus allows one to deduce the derivatives of the 
eigen values of the neutralino mass matrix analytically even though 
the eigen values themselves cannot be gotten analytically in a
compact form. We use this calculus to obtain analytical results 
for the neutralino-Z-neutral Higgs exchange contribution
to the masses and mixings in the CP even-CP odd neutral Higgs sector.
The above computation along with the stop-top, the sbottom-bottom,
the tau-stau and 
the chargino-W-
charged Higgs exchange contribution computed previously provide us with a 
complete  one loop contribution to the Higgs mass matrix with the inclusion
of CP phases. This full one
 loop result was then used to discuss the phenomenon of 
CP violation in the neutral Higgs sector. The numerical analysis 
shows that the mixings
between the CP even and the CP odd sectors are significantly affected by 
the neutralino exchange contribution.
The mixing of the CP even and the CP odd Higgs sector have many 
important consequences\cite{demir,inhiggs,carena}. Thus one consequence 
is that
CP even-odd mixing affects the couplings of the Higgs bosons with
quarks and leptons and this effect can be discerned in 
Higgs searches in collider experiments. Another important implication
is that the CP even-odd mixing will affect the relic density analysis 
and thus modify the parameter space allowed by the relic
density constraints. Further, since the couplings of the quark
and leptons with the Higgs are affected due to the CP even-odd mixing
there will also be an effect of these mixings on detection rates in
the direct searches for dark matter. It would be interesting to carry
out an analysis of these phenomena.

\noindent
{\bf Acknowledgments}\\ 
This work was initiated during the period when one of the authors
(PN) was at the Physics Institute at the University of Bonn,  
the Max-Planck Institute fuer Kernphysik, Heidelberg and CERN. 
The author acknowledges hospitality during the period of his stay and 
support from an Alexander von Humboldt award.
 This research was also supported in part by NSF grant PHY-9901057\\

\section{Appendix A: Neutralino eigen values and derivatives}
The  characteristic equation for the square of
the neutralino mass is $F(\lambda)$= 
$Det(M_{\chi^0}^{\dagger}$
$M_{\chi^0}$$-\lambda I)=0$ where $\lambda$ represents
the square of the neutralino mass eigen values. It can be  expanded as

\beq
F(\lambda)=\lambda^4+a \lambda^3+b\lambda^2+c\lambda +d=0
\eeq
In the above a,b,c and d are  computed using Eq.(14).
The computation of the co-efficients is done to leading
and to next to the leading order in an expansion in $M_Z^2/M_S^2$
where $M_S$ stands for the soft  SUSY parameters. Thus, e.g., a is
expanded to $O(M_S^2)$ and $O(M_Z^2)$ orders (it is actually
exact when expanded to this order), b is expanded to 
$O(M_S^4)$ and $O(M_S^2M_Z^2)$ orders etc. The analysis for a, b and c
(d does not enter in Eqs.(23-29) and is not exhibited) gives
\begin{eqnarray}
a=-[|\tilde m_1|^2+|\tilde m_2|^2 +2|\mu|^2 +2M_Z^2]
\end{eqnarray}
\begin{eqnarray}
b=|\tilde m_1|^2 |\tilde m_2|^2 +|\mu|^4 +2|\mu|^2(|\tilde m_1|^2+ |\tilde m_2|^2)\nonumber\\
+M_Z^2[|m_1|^2+|m_2|^2+2|\mu|^2+(|\tilde m_1|^2- |\tilde m_2|^2)\cos2\theta_W\nonumber\\
-4\cos\beta \sin\beta C^2_W |\tilde m_2||\mu|\cos \gamma_2
-4\cos\beta \sin\beta S^2_W |\tilde m_1||\mu|\cos \gamma_1]
\end{eqnarray}
where $C_W^2=\frac{g_2^2}{g_1^2+g_2^2}$ and $S_W^2=\frac{g_1^2}{g_1^2+g_2^2}$
\begin{eqnarray}
c=-2|\mu|^2|\tilde m_1|^2 |\tilde m_2|^2-
|\mu|^4(|\tilde m_1|^2+|\tilde m_2|^2)\nonumber\\
+4M_Z^2\sin\beta\cos\beta |\mu|[(|\tilde m_2|^2+|\mu|^2)S_W^2|\tilde m_1|
\cos\gamma_1+ (|\tilde m_1|^2+
|\mu|^2) C_W^2 |\tilde m_2|\cos\gamma_2]\nonumber\\
\end{eqnarray}
The derivatives $\partial\lambda_{i}/\partial \Phi_{\alpha}$ can be gotten 
explicitly as follows:
\beqn
\frac{\partial\lambda_{i}}{\partial \Phi_{\alpha}}
=-\frac{a_{\alpha}\lambda^3+b_{\alpha}\lambda^2+c_{\alpha}\lambda+d_{\alpha}}
{4\lambda^3+3a \lambda^2+2b\lambda+c}|_{\lambda =\lambda_i}
\eeqn
The second derivatives are given by
\beqn
\frac{\partial^2\lambda_i}{\partial \Phi_{\alpha}\partial \Phi_{\beta}}
=[-\frac{(a_{\alpha}\lambda^3+b_{\alpha}\lambda^2+c_{\alpha}\lambda+d_{\alpha})}
{(4\lambda^3+3a \lambda^2+2b\lambda+c)^3}
(a_{\beta}\lambda^3+b_{\beta}\lambda^2+c_{\beta}\lambda+d_{\beta})
(12\lambda^2+6a\lambda+2b)\nonumber\\
+\frac{(a_{\alpha}\lambda^3+b_{\alpha}\lambda^2+c_{\alpha}\lambda+d_{\alpha})}
{(4\lambda^3+3a \lambda^2+2b\lambda+c)^2}
(3a_{\beta}\lambda^2+2b_{\beta}\lambda+c_{\beta})
+\frac{(a_{\beta}\lambda^3+b_{\beta}\lambda^2+c_{\beta}\lambda+d_{\beta})}
{(4\lambda^3+3a \lambda^2+2b\lambda+c)^2}\nonumber\\
\times (3a_{\alpha}\lambda^2+2b_{\alpha}\lambda+c_{\alpha})
-\frac{(a_{\alpha\beta}\lambda^3+b_{\alpha\beta}\lambda^2+
c_{\alpha\beta}\lambda+d_{\alpha\beta})}
{(4\lambda^3+3a \lambda^2+2b\lambda+c)}]_{\lambda =\lambda_i}
\eeqn
where 

\beqn
a_{\alpha}=\frac{\partial a}{\partial \Phi_{\alpha}}, ~~~
a_{\alpha\beta}=\frac{\partial^2 a}{\partial \Phi_{\alpha}\partial\Phi_{\beta}}
\eeqn

\section{Appendix B: List of parameters}
The explicit evaluation of the co-efficients $a_1,b_1,c_1,d_1$ 
is given  below

\beqn
a_1=-(g^2_1+g^2_2)v_1\nonumber\\
b_1=-g^2_2 |\mu||\tilde m_2|v_2 \cos\gamma_2 
-g^2_1 |\mu||\tilde m_1|v_2 \cos\gamma_1\nonumber\\
+v_1[|\tilde m_1|^2 g^2_2 +|\tilde m_2|^2 g^2_1 +(g^2_1+g^2_2)|\mu|^2] 
\eeqn

\beqn
c_1=g^2_2 (|\tilde m_1|^2 +|\mu|^2) |\mu||\tilde m_2| v_2 \cos\gamma_2\nonumber\\
+g^2_1 (|\tilde m_2|^2 +|\mu|^2) |\mu||\tilde m_1| v_2 \cos\gamma_1\nonumber\\
-g^2_2 |\mu|^2  |\tilde m_1|^2 v_1
-g^2_1 |\mu|^2  |\tilde m_2|^2 v_1
\eeqn

\beqn
d_1=-g^2_2 |\mu|^3 |\tilde m_1|^2 |\tilde m_2| v_2 \cos\gamma_2\nonumber\\
-g^2_1 |\mu|^3 |\tilde m_2|^2 |\tilde m_1| v_2 \cos\gamma_1
\eeqn
The co-efficients $a_2, b_2, c_2, d_2$ can be gotten from $a_1, b_1, c_1, d_1$
with the following interchanges

\beqn
a_2=a_1(v_1\leftarrow \rightarrow v_2),~~~
b_2=b_1(v_1\leftarrow \rightarrow v_2),~~~ 
c_2=c_1(v_1\leftarrow \rightarrow v_2),~~~
d_2=d_1(v_1\leftarrow \rightarrow v_2)
\eeqn
 The co-efficients $a_3, b_3, c_3, d_3$ are given as follows

\beqn
a_3=0\nonumber\\
b_3=-g_2^2|\tilde m_2||\mu|v_2 \sin\gamma_2
- g_1^2 |\tilde m_1||\mu|v_2 \sin\gamma_1
\eeqn

\beqn
c_3= g_2^2 (|\tilde m_1^2|^2 +|\mu|^2) |\tilde m_2||\mu|v_2\sin\gamma_2 
+g_1^2 (|\tilde m_2^2|^2 +|\mu|^2) |\tilde m_1||\mu|v_2\sin\gamma_1 
\eeqn

\beqn
d_3=-g_2^2|\tilde m_1|^2|\mu|^3  |\tilde m_2|v_2 \sin\gamma_2
- g_1^2 |\tilde m_2|^2|\mu|^3 |\tilde m_1|v_2 \sin\gamma_1
\eeqn
The co-efficients $a_3', b_3', c_3', d_3'$ can be gotten from 
$a_3, b_3, c_3, d_3$
with the following interchanges

\beqn
a_3'=a_3(v_1\leftarrow \rightarrow v_2),~~~ 
b_3'=b_3(v_1\leftarrow \rightarrow v_2),~~~ 
c_3'=c_3(v_1\leftarrow \rightarrow v_2),~~~
d_3'=d_3(v_1\leftarrow \rightarrow v_2)
\eeqn
$A_0$ and $B_0$ are given by

\beqn
A_0=2(g_1^2+g_2^2)v_1 (M_Z^2-M_{A^0}^2)\cos 2\beta
+(g_1^2+g_2^2)v_2 (M_Z^2+M_{A^0}^2)\sin 2\beta
\eeqn

\beqn
B_0=-2(g_1^2+g_2^2)v_2 (M_Z^2-M_{A^0}^2)\cos 2\beta
+(g_1^2+g_2^2)v_1 (M_Z^2+M_{A^0}^2)\sin 2\beta
\eeqn

\newpage
\noindent
{\bf Figure Captions}\\
Fig.1: Plot of the CP even component $\phi_1$ of $H_1$ (upper 
curves)  and the CP odd component $\psi_{1D}$ of $H_1$ (lower curves)
including the stop, sbottom, stau, chargino and neutralino sector contributions 
as a function of the scale $Q$.
The common parameters are  $m_A=300$, $\tan\beta=15$,
$m_0=100$, $m_{\frac{1}{2}}=500$, $\xi_1=.4$, $\xi_2=.5$, $\alpha_0=.3$, $|A_0|=1$.
The curves with circles are for $\theta_{\mu} =0.1$ and with 
 squares for $\theta_{\mu} =0.2$ where all masses are in GeV
and all angles are in radians.\\

\noindent
Fig.2: Plot of $\Delta_{13}$ 
including the stop, sbottom, stau, chargino and neutralino sector contributions 
vs the U(1) gaugino phase $\xi_1$. 
The common input for all  the curves are 
$m_0=100$,  $m_{\frac{1}{2}}=500$,
$M_A=300$, $|A_0|=1$, $\alpha_0=0.3$, 
$\xi_2=0.5$ and $Q=320$. 
The five curves correspond to the pairs
 of $\tan\beta$ and $\theta_{\mu}$ values as follows. The curve with 
 $\Delta_{13}= 301$ at $\xi_1=0$ corresponds to $\tan\beta=5$, 
 $\theta_{\mu} = .4$. Similarly the curves with values of 
 $\Delta_{13}=406$ at $\xi_1=0$ correspond to  $\tan\beta=6$, $\theta_{\mu} = .6$,
 $\Delta_{13}=416$ at $\xi_1=0$ correspond to  $\tan\beta=10$, $\theta_{\mu} = .2$,
 $\Delta_{13}=501$ at $\xi_1=0$ correspond to  $\tan\beta=8$, $\theta_{\mu} = .8$,
 and 
 $\Delta_{13}=579$ at $\xi_1=0$ correspond to  $\tan\beta=15$, $\theta_{\mu} = .3$
where all masses are in GeV and all angles are in radians. \\

\noindent
Fig.3: Plot of $\Delta_{23}$ 
including the stop, sbottom, stau, chargino and neutralino sector contributions 
vs the U(1) gaugino phase $\xi_1$ for the same input parameters as in
Fig.2. The curves with the same symbols  as in Fig.2  have the same
common inputs. \\

\noindent
Fig.4: Plot of the CP even component $\phi_1$ of $H_1$ (upper curves)
and the  CP odd component $\psi_{1D}$ of $H_1$ (lower curves)
including the stop, sbottom, stau, chargino and neutralino sector contributions 
  as a function of the U(1) gaugino phase  $\xi_1$  for the same inputs as 
 in Fig.2. The curves with the same symbols as in Fig.2 have the same
common inputs. \\ 

\noindent
Fig.5: Plot of the CP even component $\phi_1$ of $H_1$ (upper 
curves)  and the CP odd component $\psi_{1D}$ of $H_1$ (lower curves)
including the stop, sbottom, stau, chargino and neutralino sector contributions 
as a function of the $\xi_2$.
The common parameters are: $m_A=300$, $Q=320$,
$m_0=100$, $m_{\frac{1}{2}}=500$, $\alpha_0=.3$, $|A_0|=1$, 
$\theta_{\mu}=.4$. For the curves with diamonds $\tan\beta=15$, $\xi_1=1.5$,
for squares $\tan\beta=8$, $\xi_1=1.5$, 
for triangles $\tan\beta=8$, $\xi_1=0.5$, and 
for circles $\tan\beta=10$, $\xi_1=1.5$ where all masses are in GeV
and all angles are in radians.\\

\noindent
Fig.6: Plot of the CP even component $\phi_1$ of $H_1$ (upper 
curves)  and the CP odd component $\psi_{1D}$ of $H_1$ (lower curves)
including the stop, sbottom, stau, chargino and neutralino sector contributions 
as a function of $\theta_{\mu}$.
The common parameters are  $m_A=300$, $Q=320$,
$m_0=100$, $m_{\frac{1}{2}}=500$, $\xi_2=.5$, $\alpha_0=.3$, $|A_0|=1$.
For curves with diamonds $\tan\beta=15$, $\xi_1=1.5$,
for squares $\tan\beta=8$, $\xi_1=1.5$, and 
for triangles $\tan\beta=8$, $\xi_1=0.5$  where all masses are in GeV
and all angles are in radians.\\

\noindent
Fig.7: Plot of the CP even component $\phi_1$ of $H_1$ (upper 
curves)  and the CP odd component $\psi_{1D}$ of $H_1$ (lower curves)
including the stop, sbottom, stau, chargino and neutralino sector contributions 
as a function of $\tan\beta$. The common input parameters for the curves
are $m_A=300$, $Q=320$, $m_0=100$, $m_{\frac{1}{2}}=500$,
$\xi_1=.5$, $\xi_2=.5$, $\alpha_0=.3$, and $|A_0|=1$.
For the curves with diamonds, $\theta_{\mu}=.4$, 
for squares $\theta_{\mu}=.6$, and for triangles $\theta_{\mu}=.8$
where all masses are in GeV and all angles are in radians.


\newpage

\begin{thebibliography}{99}
\bibitem{ellis} 
See, e.g., J. Ellis, S. Ferrara and D.V. Nanopoulos, 
Phys. Lett. {\bf B114}, 231(1982);
 J. Polchinski and M.B. Wise, Phys.Lett.{\bf B125},393(1983);
  E. Franco and M. Mangano, Phys.Lett.{\bf B135},445(1984); 
 R.Garisto and J. Wells, Phys. Rev. {\bf D55}, 611(1997);
M. Dugan, B. Grinstein and L. Hall, Nucl. Phys. {\bf B255},
413(1985).


\bibitem{na} 
P. Nath, Phys. Rev. Lett.{\bf 66}, 2565(1991); 
Y. Kizukuri and  N. Oshimo, Phys.Rev.{\bf D46},3025(1992.




\bibitem{edmexp}
E. Commins, et. al., Phys. Rev. {\bf A50}, 2960(1994);
K. Abdullah, et. al., Phys. Rev. Lett. {\bf 65}, 234(1990),
P.G. Harris et.al., Phys. Rev. Lett. {\bf 82}, 904(1999).



\bibitem{in1}
 T. Ibrahim and P. Nath, Phys. Lett. {\bf B 418}, 98(1998);
T. Ibrahim and P. Nath,
 Phys. Rev. {\bf D57}, 478(1998); E ibid {\bf D58}, 019901(1998); 
   Phys. Rev. {\bf D58}, 111301(1998);
 T. Falk and K Olive, Phys. Lett. {\bf B 439}, 71(1998);
 M. Brhlik, G.J. Good, and G.L. Kane, Phys. Rev. {\bf D59}, 115004
 (1999); A. Bartl, T. Gajdosik, W. Porod, P. Stockinger, and
 H. Stremnitzer,  Phys. Rev. {\bf 60}, 073003(1999);
 T. Falk, K.A. Olive, M. Prospelov, and R. Roiban, Nucl. Phys. 
 {\bf B560}, 3(1999); 
 S. Pokorski, J. Rosiek and C.A. Savoy, 
 Nucl. Phys. {\bf B570}, 81(2000);
 M. Brhlik, L. Everett, G. Kane and J. Lykken, Phys. Rev.
 Lett. {\bf 83}, 2124, 1999; 
 hep-ph/9908326; E. Accomando, R. Arnowitt and B. Datta, 
  hep-ph/9907446; Phys. Rev. {\bf D61}, 
 075010(2000);
T. Ibrahim and P. Nath, Phys. Rev. {\bf D61}, 093004(2000);
S.Abel, S. Khalil, O.Lebedev, Phys. Rev. Lett. {\bf 86}, 5850(2001);
 U. Chattopadhyay, T. Ibrahim, D.P. Roy,Phys. Rev. {\bf{D64}}:013004,2001. 



\bibitem{chang}
D. Chang, W-Y.Keung,and A. Pilaftsis, Phys. Rev. Lett. {\bf 82}, 
900(1999). 



\bibitem{bdm1}
K.S. Babu, B. Dutta and R. N. Mohapatra, Phys. Rev. {\bf D61}, 
091701(2000).


\bibitem{ccn}
K.L. Chan, U. Chattopadhyay and P. Nath, Phys. Rev. {\bf D58}, 
096004(1998).

\bibitem{ing2}
T. Ibrahim and P. Nath, Phys. Rev. {\bf D61}, 095008(2000);
T. Ibrahim and P. Nath, Phys. Rev. {\bf D62}, 015004(2000);
hep-ph/9908443: 
T. Ibrahim, U. Chattopadhyay and P. Nath, Phys.\ Rev.\ D {\bf 64}, 
016010(2001).


\bibitem{olive}
T. Falk, K.A. Olive and M. Srednicki, Phys. Lett. 
{\bf B354}, 99(1995);
U. Chattopadhyay, T. Ibrahim and P. Nath, Phys. Rev. {\bf D60}, 
063505(1999); hep-ph/0005109; 
T. Falk, A. Ferstl and K. Olive, hep-ph/9908311;
S. Khalil and Q. Shafi, Nucl. Phys. {\bf B564},19(1999). K. Freese 
amd P.  Gondolo, hep-ph/9908390; S.Y. Choi, hep-ph/9908397;
S. Khalil,   Phys. Lett. {\bf B484}, 98(2000).

\bibitem{trilep}
P. Nath and R. Arnowitt, Mod. Phys.Lett.{\bf A2}, 331(1987);
H. Baer and X. Tata, Phys. Rev.{\bf D47}, 2739(1993);
S.Y. Choi, H.S. Song, and W.Y. Song, Phys. Rev.{\bf B483}, 
168(2000); hep-ph/0007276;  S.Y. Choi, M. Guchait, H.S. Song, and W.Y. Song, 
 Phys.Lett.{\bf B483}, 168(2000).


\bibitem{baryogenesis}
M. Carena, J.M. Moreno, M. Quiros, M. Seco, 
C.E.M. Wagner, Nucl.Phys.{\bf B599},158(2001). 

\bibitem{other}
T. Ibrahim and P. Nath, Phys. Rev. {\bf{D62}}:095001,2000; hep-ph/0004098;
T. Ibrahim, Phys.Rev.{\bf D64}, 035009(2001);
V. Barger, Tao Han, Tian-Jun Li, Tilman Plehn,
 Phys.Lett.B475:342-350,2000;  V. Barger, T. Falk, T. Han, 
 J. Jiang, T. Li, T. Plehn, hep-ph/0101106;
S. Mrenna, G.L.Kane and L-T Wang, Phys. Lett.{\bf B483}, 175(2000);
M. Brhlik, L. Everett, G.L. Kane, S.F. King, and O. Lebedev, 
Phys. Rev. Lett. {\bf 84}, 3041(2000);  J.M. Frere and M. Gavela,
Phys. Lett.{\bf B132}, 107(1983);
O.C.W. Kong, Nucl. Phys. Proc. Suppl.{\bf 101}, 421(2001);
A. Dedes, S. Moretti, Phys.Rev.Lett.84:22-25,2000;
 Nucl.Phys.B576:29-55,2000;  S.Y. Choi, J.S. Lee, Phys. Rev. {\bf{D61}}:015003,2000;
D.A. Demir and M.B. Voloshin, Phys. Rev. {\bf{D63}}:115011,2000;
C.~S.~Huang and W.~Liao,
Phys.\ Rev.\ D {\bf 62}, 016008 (2000);Phys.\ Rev.\ D {\bf 61}, 116002 (2000);
A.G. Akeroyd and A. Arhrib, hep-ph/0107040;
T. Ibrahim and P. Nath, Phys. Rev. {\bf D64}, 093002(2001);
hep-ph/0107325;
J.L. Feng, K.T. Matchev, and Y. Shadmi, Nucl Phys. {\bf B613}, 366(2001);
hep-ph/01017182; A. Bartl, K. Hidaka, T. Kernreiter and  W. Porod,
hep-ph/0204071. 




\bibitem{pilaftsis}
A. Pilaftsis, Phys. Rev. {\bf D58}, 096010; Phys. Lett.{\bf B435}, 
88(1998).

\bibitem{pilaftsis2}
A. Pilaftsis and C.E.M. Wagner, Nucl. Phys. {\bf B553}, 3(1999).

\bibitem{demir}
D.A. Demir, Phys. Rev. {\bf D60}, 055006(1999);
Mugo Boz, hep-ph/0008052.


\bibitem{inhiggs}
T. Ibrahim and P. Nath,  
Phys.Rev. {\bf D63}, 035009(2001).


\bibitem{drees}
C.Y.Choi, M. Drees and J.S. Lee, Phys. Lett. {\bf B481}, 57(2000).

\bibitem{carena}
M. Carena, J. Ellis, A. Pilaftsis and C.E.M. Wagner,
Nucl. Phys. {\bf B586}, 92(2000);  Nucl.Phys.{\bf B625},
345(2002); hep-ph/0003180; hep-ph/0111245 




\bibitem{ellis0}
J. Ellis, G. Ridolfi and F. Zwirner, Phys. Lett. {\bf B257}, 83(1991);
M.S. Berger,Phys. Rev. {\bf D41}, 225(1990); 
H.E. Haber and R. Hempling,Phys. Rev. Lett. {\bf 66}, 1815(1991);
Y. Okada, M. Yamaguchi and T. Yanagida, Prog. Theor. Phys. {\bf 85},
 1(1991); 
R. Barbieri, M. Frigeni and F, Caravaglio, Phys. Lett. {\bf B258},
167(1991);
A. Brignole, J.Ellis, G. Ridolfi and F. Zwirner, Phys. Lett.
{\bf 371}, 123(1991); 
M. Drees and M. Nojiri, Phys. Rev.{\bf D45}, 2482(1992);
P.H. Chankowski, S. Pokorski and J. Rosiek, Phys. Lett. {\bf B281},
100(1992); Nucl. Phys.{\bf B423}, 437(1994);  D.M. Pierce,
A. Papadopoulos and S.B. Johnson, Phys. Rev.Lett. {\bf 68}, 
3678(1992); A. Brignole, Phys. Lett.{\bf B281}, 284(1992);
M. Drees and M.M. Nojiri, Phys. Rev. {\bf D45}, 2482(1992); 
V. Barger, M.S. Berger and P. Ohmann, Phys. Rev.{\bf D49}, 4908(1994).
R. Hempfling and A.H. Hoang, Phys. Lett.{\bf B33}, 99(1994);
 Haber and R. Hempfling, Phys. Rev. {\bf D48}, 4280(1993);
J. Kodaira, Y. Yasui and K. Sasaki, Phys. Rev. {\bf D50}, 7035(1994);
J.A. Casas, J.R. Espinosa, M. Quiros and A. Riotto,
Nucl. Phys.{\bf B436}, 3(1995); M. Carena, J.R. Espinosa, M. Quiros
and C. Wagner, Phys. Lett. {\bf B355}, 209(1995);
 M. Carena,  M. Quiros and C. Wagner, Nucl. Phys. {\bf B461}, 407(1996);
 A.V. Gladyshev, D.I. Kazakov, W. de Boer, G. Burkart, R. Ehret,
 Nucl. Phys. {\bf B498}, 3(1997);
S. Heinemeyer, W. Hollik and QG. Weiglein, 
Phys. Lett. {\bf B440}, 96(1998); Phys. Rev. {\bf D58}, 091701(1998);
R.-J. Zhang, Phys. Lett. {\bf B447}, 89(1999); R.-J. Zhang,
JHEP, {\bf 0003}, 026(2000); J.R. Espinosa and R,J. Zhang, Nucl. Phys.
{\bf B586}, 3(2000). 
%hep-ph/9912236.

\bibitem{applied}
A.H. Chamseddine, R. Arnowitt and P. Nath, \Journal{\PRL}{49}
{970}{1982}; R. Barbieri, S. Ferrara and C.A. Savoy, \Journal{\PLB}
{119}{343}{1982}; L. Hall, J. Lykken, and S. Weinberg,
\Journal{\PRD}{27}{2359}{1983}: P. Nath, R. Arnowitt and A.H. Chamseddine,
\Journal{\NPB}{227}{121}{1983}. For reviews, see P. Nath, R. Arnowitt
and A.H. Chamseddine, "Applied N=1 Supergravity", world scientific,
1984; H.P. Nilles, Phys. Rep. {\bf 110}, 1(1984).





\bibitem{coleman}
 S. Coleman and E. Weinberg, Phys. Rev. {\bf D7}, 1888(1973); 
S. Weinberg, ibid., {\bf 7}, 2887(1973).


\bibitem{arnowitt1} 
R. Arnowitt and P. Nath, Phys. Rev. {\bf D46}, 3981(1992).


\end{thebibliography}
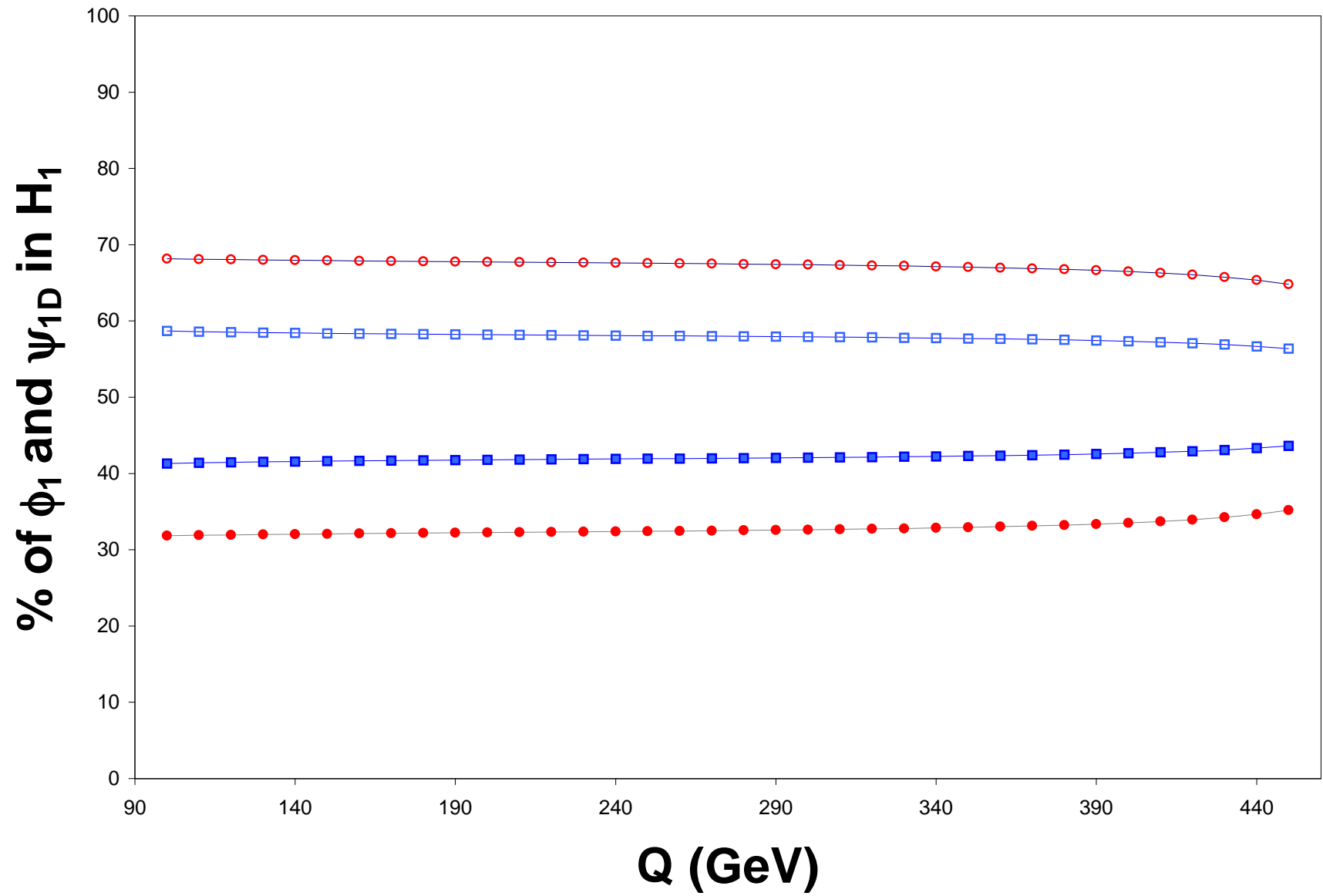
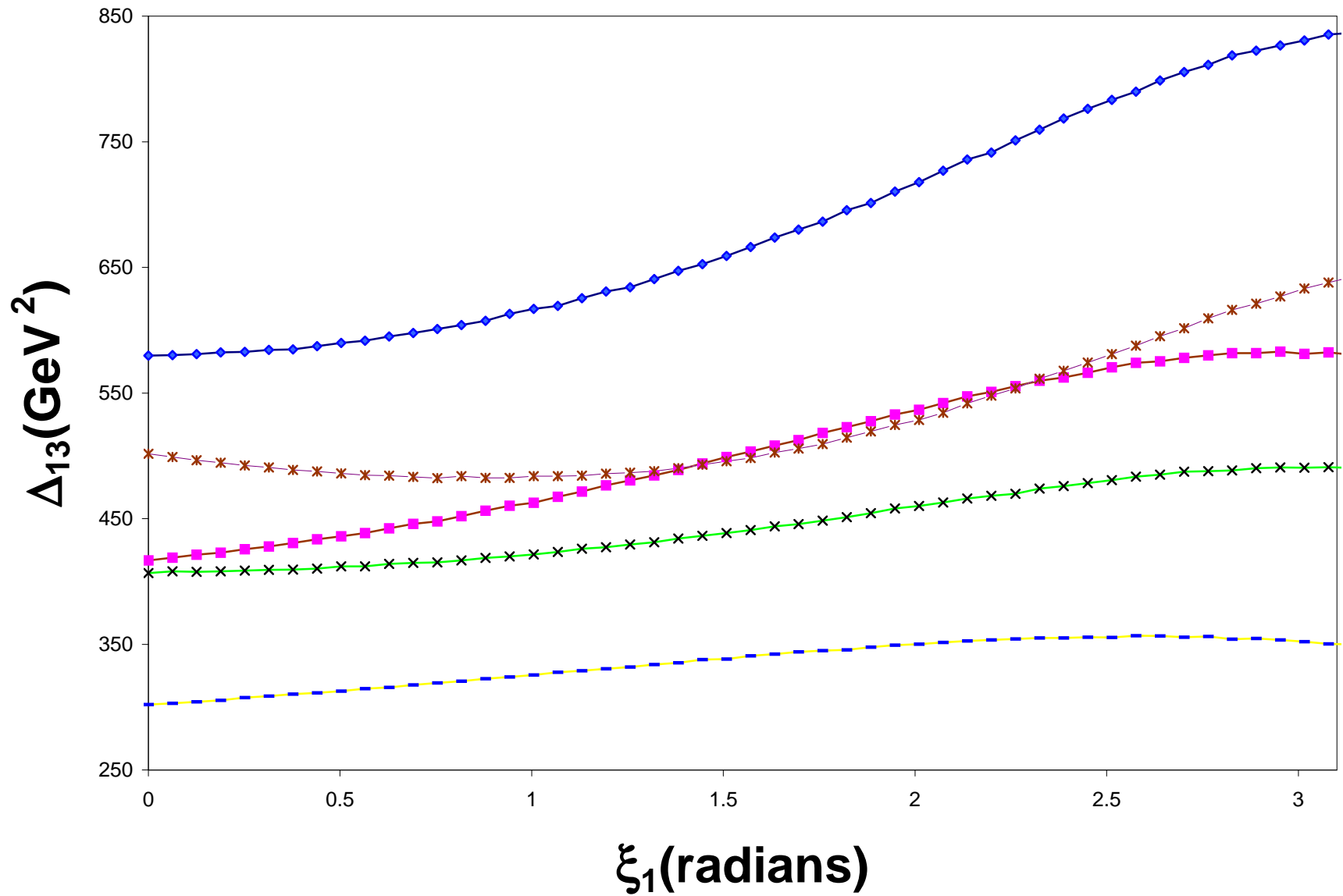
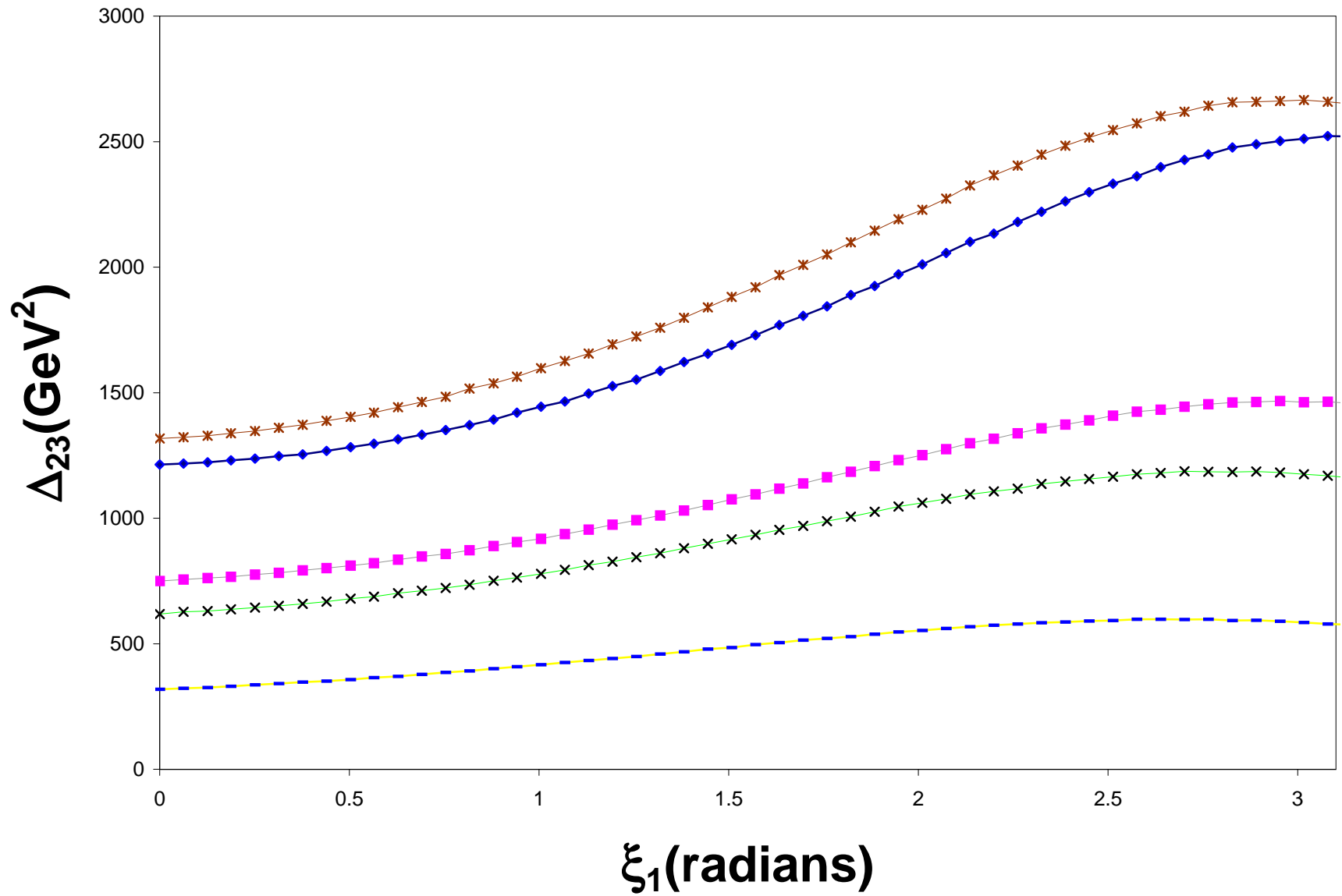
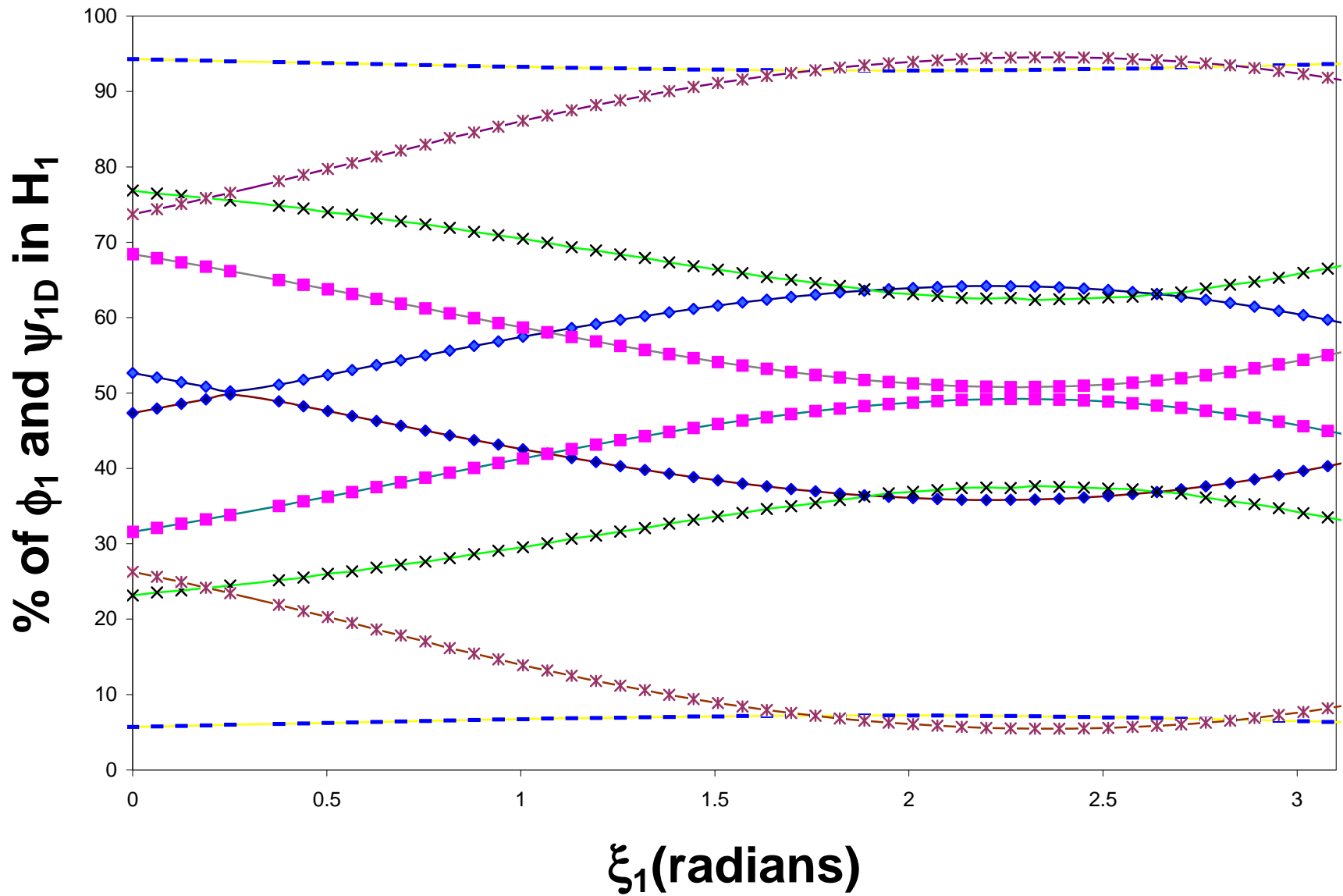
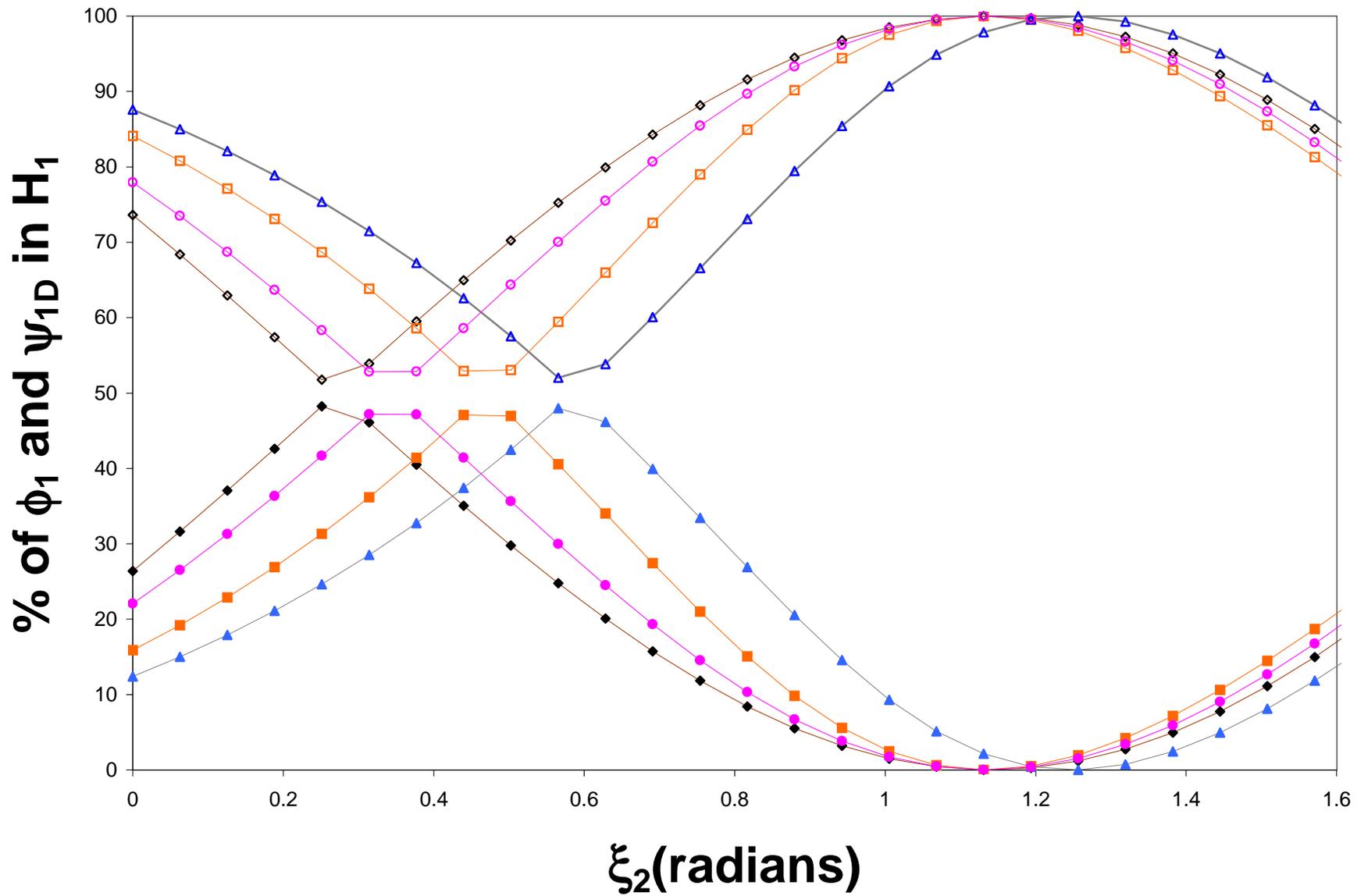
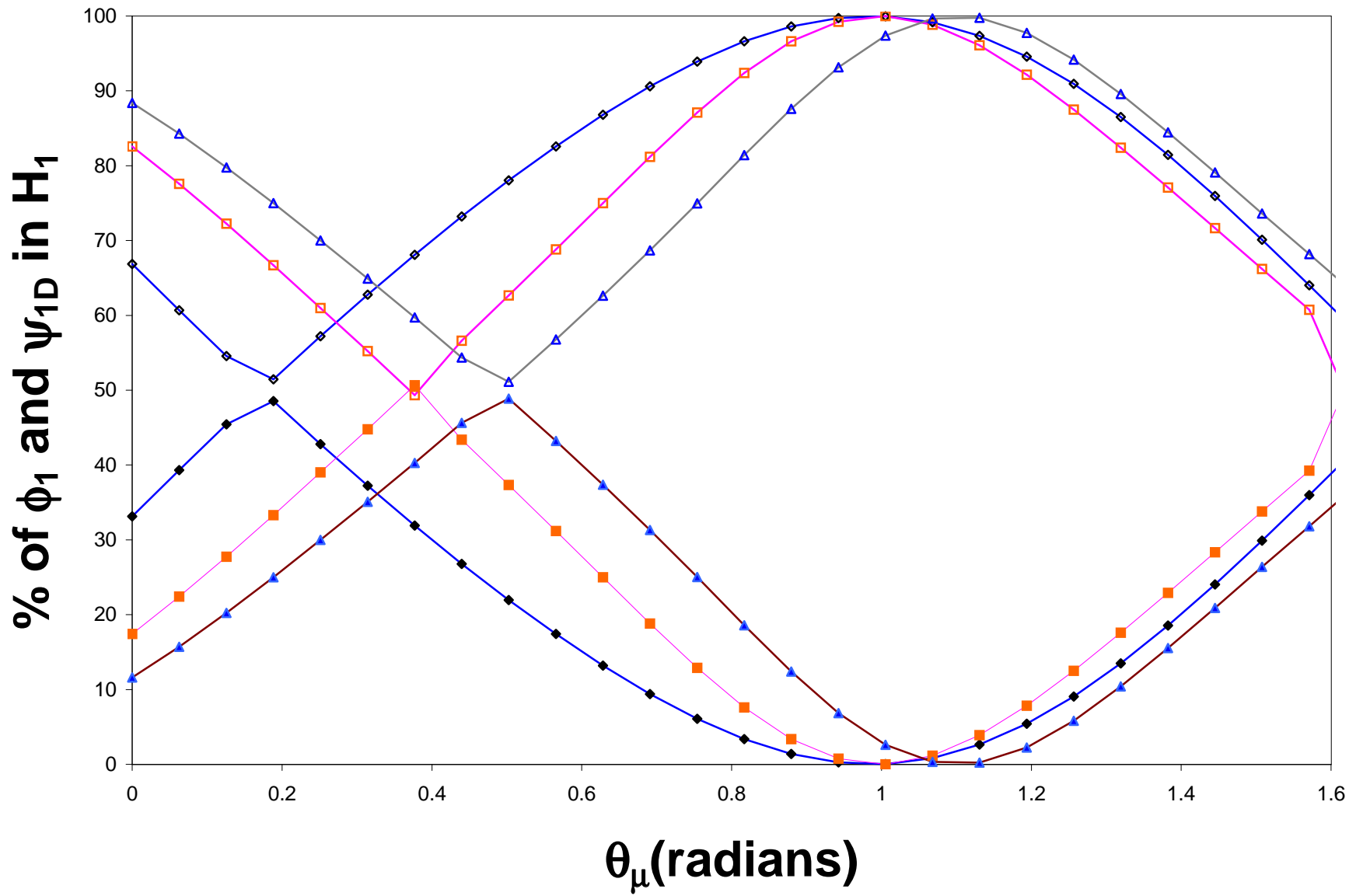
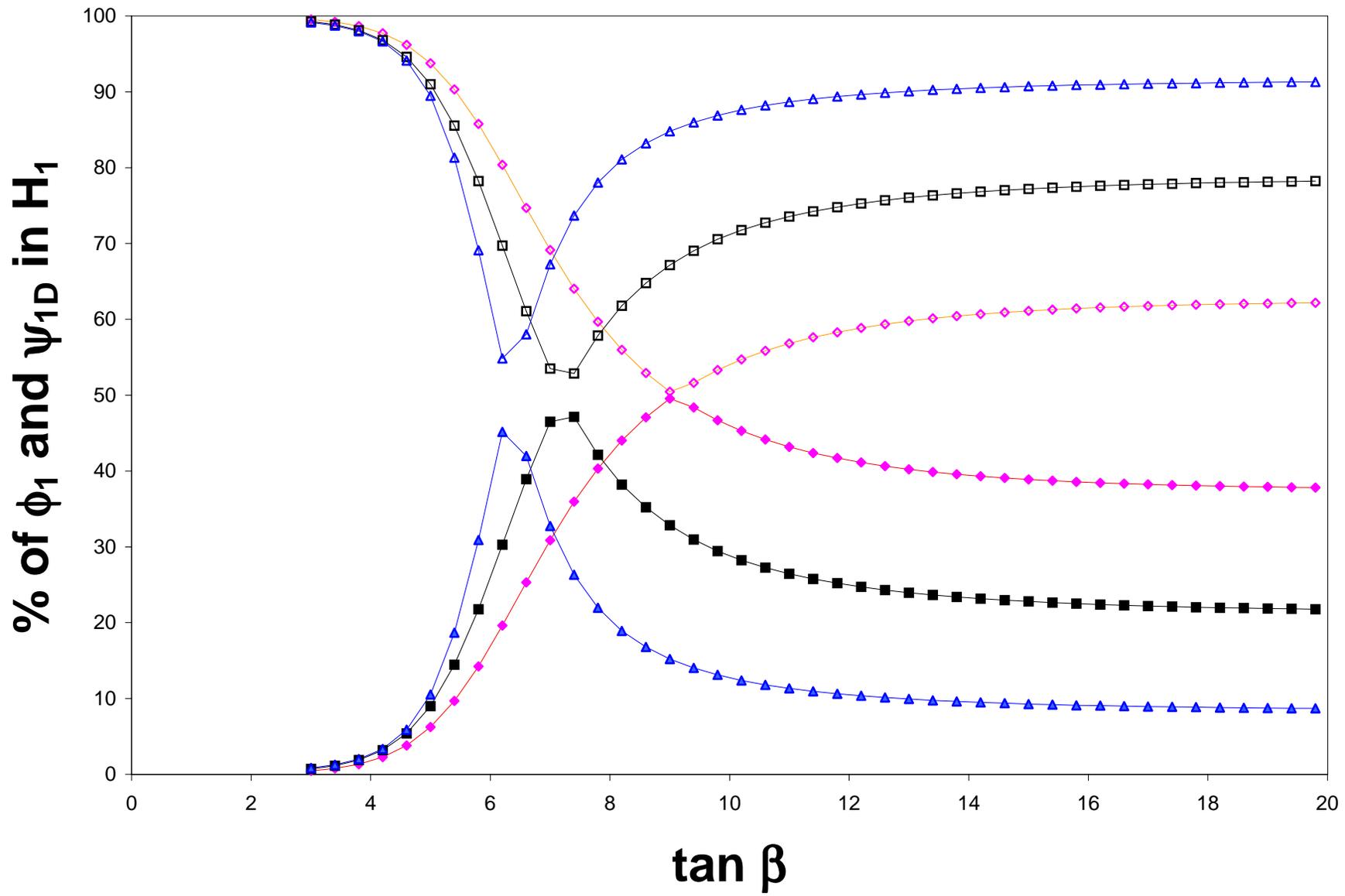
\end{document}